\documentclass[aip,apl,amsmath,amssymb,reprint]{revtex4-1}
\usepackage[T1]{fontenc} 
\usepackage{graphicx}
\usepackage{dcolumn}
\usepackage{bm}
\usepackage{comment}
\usepackage{soul}
\usepackage[dvipsnames]{xcolor}
\usepackage{color}

\begin{document}

\preprint{AIP/123-QED}

\title{Relativistic surface-plasmon enhanced harmonic generation from gratings}%

\author{L.Fedeli}
\email{luca.fedeli@polimi.it} 
\affiliation{Department of Energy, Politecnico di Milano, 20133 Milano, Italy}
\affiliation{National Institute of Optics, National Research Council (CNR/INO)  A.Gozzini unit , 56124 Pisa, Italy}
\affiliation{Enrico Fermi department of Physics, University of Pisa, 56127 Pisa, Italy}
\author{A.Sgattoni}
\affiliation{LULI-UPMC: Sorbonne Universites, CNRS,  \'Ecole Polytechnique, CEA, 75013 Paris, France}
\affiliation{LESIA, Observatoire de Paris, CNRS, UPMC: Sorbonne Universites, 92195 Meudon, France}
\affiliation{National Institute of Optics, National Research Council (CNR/INO)  A.Gozzini unit , 56124 Pisa, Italy}
\author{G.Cantono}
\affiliation{LIDYL, CEA, CNRS, Université Paris-Saclay, CEA Saclay, 91191 Gif-sur-Yvette, France}
\affiliation{Universit{\'e} Paris Sud, Paris, {91400 Orsay, France}}
\affiliation{National Institute of Optics, National Research Council (CNR/INO)  A.Gozzini unit , 56124 Pisa, Italy}
\affiliation{Enrico Fermi department of Physics, University of Pisa, 56127 Pisa, Italy}
\author{A.Macchi}
\affiliation{National Institute of Optics, National Research Council (CNR/INO)  A.Gozzini unit , 56124 Pisa, Italy}
\affiliation{Enrico Fermi department of Physics, University of Pisa, 56127 Pisa, Italy}

\date{\today}

\begin{abstract}
The role of relativistic surface plasmons (SPs) in high order harmonic emission from laser-irradiated grating targets has been investigated by means of particle-in-cell simulations. SP excitation drives a strong enhancement of the intensity of harmonics, particularly in the direction close to the surface tangent. 
The SP-driven enhancement overlaps with the angular separation of harmonics generated by the grating, which is beneficial for applications requiring monochromatic XUV pulses.
\end{abstract}

\pacs{Valid PACS appear here}
\maketitle
Laser-based high-order harmonic (HH) sources have allowed an unprecedented control on ultra-short electronic processes in atomic and molecular systems, supporting the development of ``attoscience''\cite{Sansone_NatPhot2011,Krausz_RevModPhys2009}. 
Several approaches have been studied in order to optimize HH generation by atomic recollision in gas targets (see e.g. Refs. \onlinecite[and references therein]{paulN03,zepfPRL07,shinerPRL09,PopmintchevS15}). However, HH generation from gas targets is ultimately limited by the onset of ionization at intensities $\lesssim 10^{15}~\textrm{W/cm}^2$.
This issue has stimulated the study of HH emission from solid targets\cite{VonDerLinde_PRA1995,Norreys_PRL1996,Tarasevitch_PRA2000,Teubner_RevModPhys2009,Thaury_JPB2010,Bierbach_NJP2012,Brugger_PRL2012,Kahaly_PRL2013,Heissler_APB2014,PhysRevLett.116.185001}  
irradiated at ultra-high intensity ($I>10^{18}~\textrm{W/cm}^2$).
{In this regime the relativistic motion of the electrons in the fields near the target surface is responsible for HHs generation\cite{Lichters_PoP1996,Baeva_PRE2006,QuereHarm} and higher laser intensities result in more intense HHs and a much more compact set-up than for gaseous targets \cite{shawJAP13}.
Moreover, a kHz repetition rate has been demonstrated using moving tape targets \cite{BorotOL11}.
In the time domain, the HH emission from solid targets has the form of a train of attosecond pulses \cite{Plaja_JOSAB_1998} and several strategies to isolate a single attosecond pulse have been investigated \cite{Wheeler_NatPhot2012,Vincenti_PRL2012,Yeung_PRL2015}. 

Attosecond duration is not a critical issue for 
applications such as XUV-litography \cite{0022-3727-37-23-005} or photoelectron spectroscopy \cite{hufner2013photoelectron} which would rather benefit from both HH intensity increase and angular separation of the HHs,
so that an effectively monochromatic XUV source is obtained (the XUV spectral region ranges approximately between the $8$-th and the $80$-th harmonic of the laser light at the wavelength $\sim 800$~nm of Ti:Sapphire systems). To this aim, the use of grating targets has been proposed \cite{Yeung_OptLett2011,Yeung_NJP2013,PanHarm2016,PhysRevE.93.053206}, so that the $m$-th harmonic is emitted for a discrete set of angles $\theta_{mn}$ depending on the period $d$ of the grating, according to the diffraction formula \cite{RzazewskiJPB00,Lavocat-Dubuis_PRE_2009}:
\begin{equation}
\label{eq:diffraction}
{n\lambda}/{md} = {\sin(\theta_{\mbox{\tiny i}}) + \sin(\theta_{mn})}
\; ,
\end{equation}
where $\lambda$ is the laser wavelength, $n$ is the diffraction order, and $\theta_{\mbox{\tiny i}}$ is the angle of incidence of the laser pulse ($\theta_{\mbox{\tiny i}}$ and $\theta_{mn}$ are referred with respect to the normal to the target as indicated in Fig.\ref{fig:bz}). This effect was experimentally observed using a laser pulse with ultra-high contrast in order to prevent early prepulse-induced damage of the grating\cite{Cerchez_PRL2013}.
The diffraction of harmonics at different angles allows to separate them from the reflected laser light which should otherwise be filtered out, being typically much more intense than the HH signal. However, the dispersion into several orders reduces the HH intensity, so that an enhancement of the HH generation efficiency is highly desirable.

Grating targets also allow the excitation of surface plasmons (SPs). Experimental evidence of SP excitation in the relativistic regime has been provided recently. 
Irradiating gratings at the resonant condition for SP excitation has been observed to both increase the cut-off energy of the ions emitted from the target\cite{Ceccotti_PRL2013} and to allow for electron acceleration along the target surface in the SP field\cite{Fedeli_PRL2016,SgattoPPCF}. 
Since the excitation of a SP is associated with a strong field enhancement at the target surface, one may expect as well a SP-enhancement of HH emission, which may overlap to the grating diffraction effects. 
Indeed, SP-enhancement of HH generation in micro and nano-structures has been previously demonstrated at low laser intensities (see e.g. Refs. \onlinecite{kimN08,HusakouOE11,parkN11,sivisNP13,hurstPRB14,hanNC16}).

In this Letter, we investigate via numerical simulations the HH emission from a grating target irradiated at the resonance condition for SP excitation. A strong enhancement of the intensity of the angularly dispersed HHs is observed near resonance. We study in detail the angular distribution of HHs, in order to individuate suitable configurations for an XUV source.

The excitation of a SP by oblique incidence of a laser pulse on a periodically modulated interface is a key process in plasmonics \cite{bookMaier2007}. In the linear regime, the matching conditions for resonant excitation of a SP by an EM plane wave at an interface between vacuum and a medium with spatial period $d$ are
\begin{equation}
\omega=\omega_{\mbox{\tiny SP}}
\; , \qquad 
\frac{\omega}{c}\sin\theta_{\mbox{\tiny res}}+ k_{\mbox{\tiny SP}} = \frac{2\pi}{d}j \; ,
\end{equation}
where $\omega=2\pi c/\lambda$ is the frequency of the incident EM wave, $\theta_{\mbox{\tiny res}}$ is the resonant angle of incidence, $j$ is an integer number, and $\omega_{\mbox{\tiny SP}}$ and $k_{\mbox{\tiny SP}}$ are the frequency and wavevector of the SP.
For a free electron metal described by the cold plasma dielectric function $\varepsilon(\omega)=1-\omega_p^2/\omega^2 \equiv 1-\alpha$ and assuming $\alpha>2$, using the SP dispersion relation (see e.g. Ref. \onlinecite[sec.2.2]{bookMaier2007} or Ref. \onlinecite[sec.68]{landau8})
$(k_{\mbox{\tiny SP}}c/\omega_{\mbox{\tiny SP}})^2=\varepsilon(\omega_{\mbox{\tiny SP}})/(\varepsilon(\omega_{\mbox{\tiny SP}})+1)$ 
we obtain
\begin{equation}\label{form:plares}
j\lambda/d = {(\alpha-1)^{1/2}/(\alpha-2)^{1/2}}+\sin(\theta_{\mbox{\tiny res}}) \; .
\end{equation}
Notice that $\alpha$ can be written also as $\alpha=n_e/n_c$ where $n_e$ is the electron density and $n_c=m_e\omega^2/4\pi e^2=1.1 \times 10^{21}~\mbox{cm}^{-3}/(\lambda[\mu\mbox{m}])^2$ is the cut-off (or critical) density for the EM wave. For a solid density material $\alpha \gg 1$, so that $\sin(\theta_{\mbox{\tiny res}}) \simeq -1+j\lambda/d$. Thus, excitation of SPs requires grating periods $d>j\lambda/2$, and does not play a role in HH generation from sub-wavelength gratings which were investigated in previous works\cite{Lavocat-Dubuis_PRE_2009,Lavocat-Dubuis_PoP_2010,Cerchez_PRL2013}. 
In the following we assume $j=1$ which corresponds to matching at the first ``band'' of the SP dispersion relation $\omega_{\mbox{\tiny SP}}(k_{\mbox{\tiny SP}})$ folded in the Brillouin zone. Notice that for $j\geq 2$ the grating period $d$ might become larger than the laser spot radius for tight focusing, which would affect the matching conditions. 
At resonance, using Eq.(\ref{eq:diffraction}) we obtain that the angles at which the $m$-th harmonic is emitted are given by
\begin{equation}
\sin(\theta_{mn}) \simeq -\sin(\theta_{\mbox{\tiny res}})+\frac{n}{m}(1+\sin(\theta_{\mbox{\tiny res}})) \; ,
\end{equation}
At high intensities, modeling the target as a cold plasma is adequate for any material because free electrons are created by ultrafast field ionization and the oscillation energy greatly exceeds the thermal energy. In such regime the coupling within the grating requires a laser pulse with ultrashort duration (few tens of fs) and very high pulse-to-prepulse contrast\cite{Dromey_RSI_2004,Levy_OptLett2007,Kapten_OptLett1991,Thaury_NatPhys2008} to prevent the grating from being washed out by hydrodynamical expansion either during or before the intense pulse. In addition, although a detailed theory of relativistic SPs is still lacking, in the relativistic regime one may expect the SP dispersion relation and, consequently, the resonance condition to be modified by nonlinear effects. However, experiments at relativistic laser intensities\cite{Ceccotti_PRL2013,Fedeli_PRL2016} have provided evidence of SP excitation at angles close to the value predicted by Eq.(\ref{form:plares}) for $\alpha\gg 1$. 
\begin{figure}
\includegraphics[width=\columnwidth]{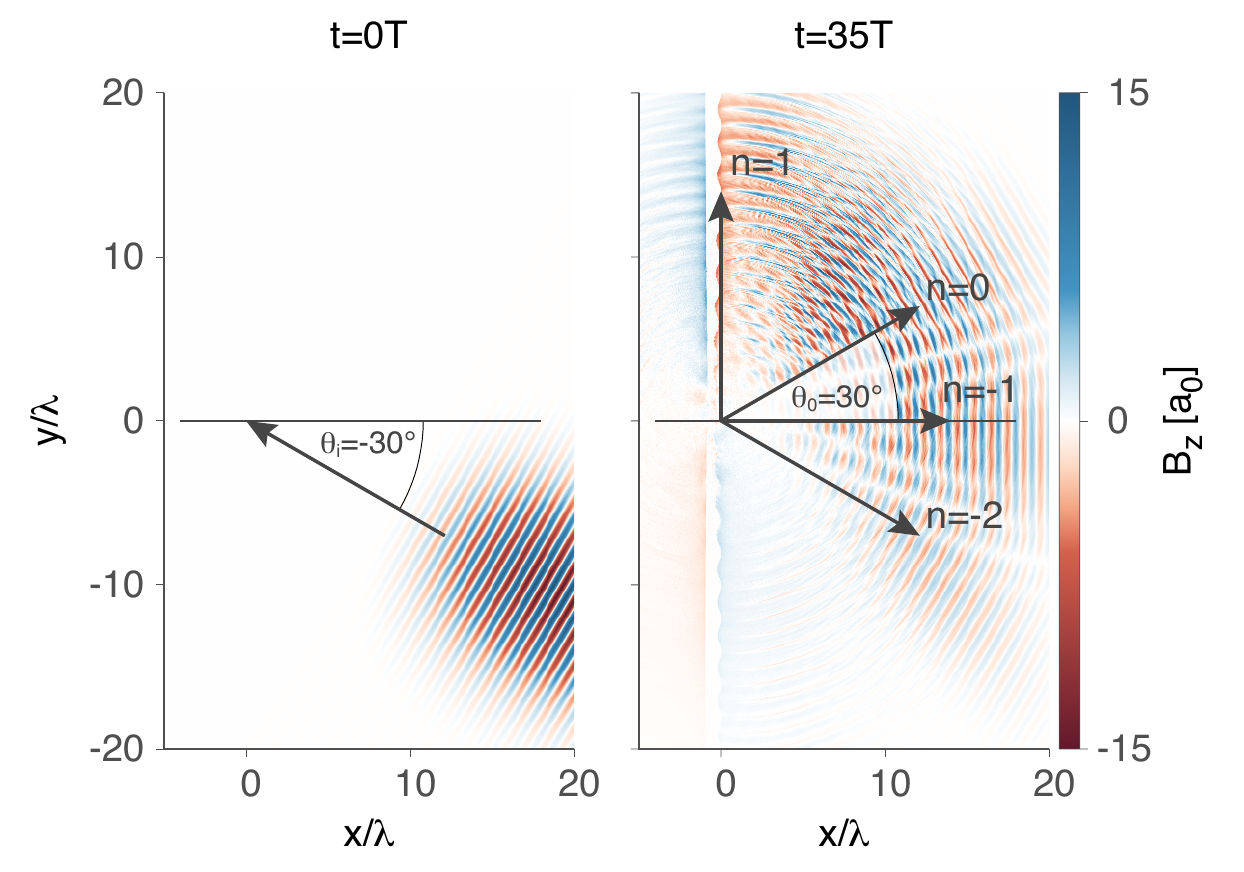}
\caption{\label{fig:bz} The $B_z$ field component for a grating target ($d=2\lambda$) irradiated at $30^{\circ}$ incidence, at times $t = 0T$ and $t = 35 T$. In addition to specular reflection at $30^{\circ}$
( $\theta_{\mbox{\tiny i} }= \theta_{\mbox{\tiny res}}=-30^{\circ}$),
diffraction of the laser pulse at the $n=1, -1,-2$ orders (corresponding to angles $90^{\circ}$, $0^{\circ}$ and $-30^{\circ}$, respectively)
and localized fields along the target surface are observed.}
\end{figure}
\begin{figure}
\includegraphics[width=1.0\columnwidth]{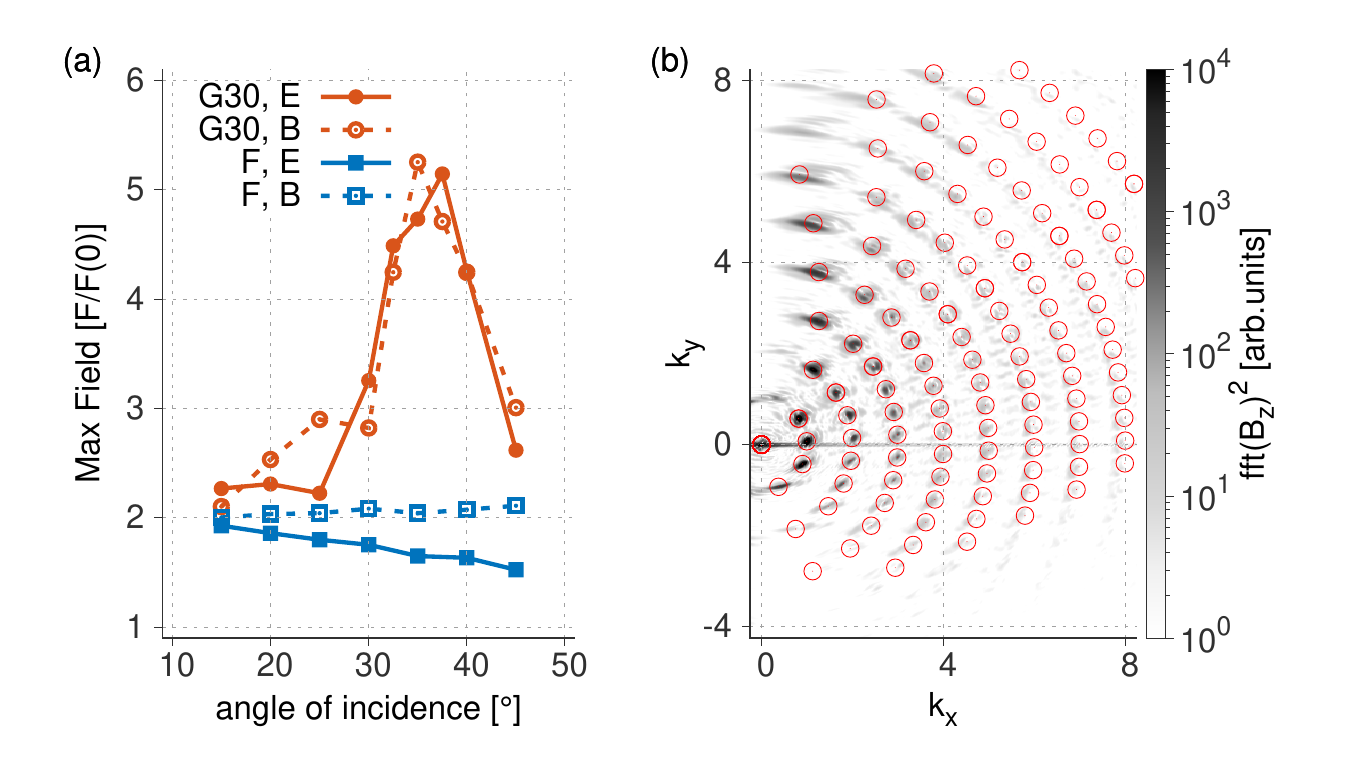}
\caption{\label{fig:largefig} a) Peak values of the electric (E) and magnetic (B) fields (normalized to the laser field initial amplitude) for flat (F) and grating (G30) targets as a function of the angle of incidence; a field enhancement peaking at $\simeq 35^{\circ}$ is observed for the grating.
b) Two-dimensional Fourier transform $\hat{B}_z(k_x,k_y)$ for $35^{\circ}$ incidence angle on a grating target. The vertical direction is along the target surface. {Red} spots show the location of the harmonic wavevectors according to Eq.(\ref{eq:diffraction}).}
\end{figure}

The simulations were performed with the open source particle-in-cell (PIC) code PICCANTE \cite{piccante,arxiv_piccante}.
The electron density was $n_e=128n_c$, which is representative of highly ionized solid targets (notice that the results do not depend either on the target density, as far as $n_e\gg n_c$, or on the target material; in a real experiment, however, a transparent material with high damage threshold to laser prepulses would be preferred). 
The target thickness was kept to $1\lambda$ for computational feasibility; however, larger values do not affect the results significantly, as expected since the fields are evanescent into the target.   
The grating period was $d=2\lambda$, corresponding to a resonance angle $\theta_{\mbox{\tiny res}}=30^{\circ}$ according to Eq.(\ref{form:plares}). The peak-to-valley depth of the grooves was $0.25 \lambda$.  
All the simulations were two-dimensional and the numerical box size was $80\lambda \times 80 \lambda$, wide enough for the boundaries not to affect the results. The number of particles per cell was 144 for each species. For most of the simulations the spatial resolution was $\lambda/154$ in each direction, in order to well resolve harmonic orders up to $m\approx 15$ with $\approx 10$ points per wavelength. Few selected cases were simulated also with increased resolution in order to resolve higher order harmonics.
The laser pulse had a Gaussian transverse profile with a waist of $5 \lambda$, a $\cos^2$ temporal profile (for the EM field) with 12$\lambda/c$ duration (FWHM), and $P$-polarization. The angle of incidence of the pulse was varied in the $5^{\circ}-45^{\circ}$ range.
The peak amplitude of the laser field in normalized units was $a_0=15$ (where $a_0=0.85(I[10^{18}\textrm{W/cm}^{2}])^{1/2}\lambda_L[\mu\textrm{m}]$) which, for $\lambda=0.8~\mu\textrm{m}$, corresponds to a peak intensity $\approx 4.9\times 10^{20}~ \textrm{W cm}^{-2}$ at the focus, presently accessible in several laser facilities \cite{HPL:9533946}. 
Additional simulations (not shown in the paper) were performed for different amplitudes in the $a_0=5 - 100$ range ($I=5.4\times 10^{19} - 2.2\times 10^{22}~\textrm{W cm}^{-2}$) and for two different values of the grating periodicity, i.e. $d=1.52 \lambda$ and $3.41 \lambda$ (corresponding to a resonance angle $\theta_{\mbox{\tiny res}}=20^{\circ}$ and $45^{\circ}$, respectively); the results were similar to those obtained for $a_0=15$ and $d=2\lambda$ so that we will restrict ourselves to this latter case in the following. The grating target and the flat target will be referred to as, respectively, G30 and F. 

\onecolumngrid
\begin{center}
\begin{figure}[h!]
\centering
\includegraphics[width=1.0\columnwidth]{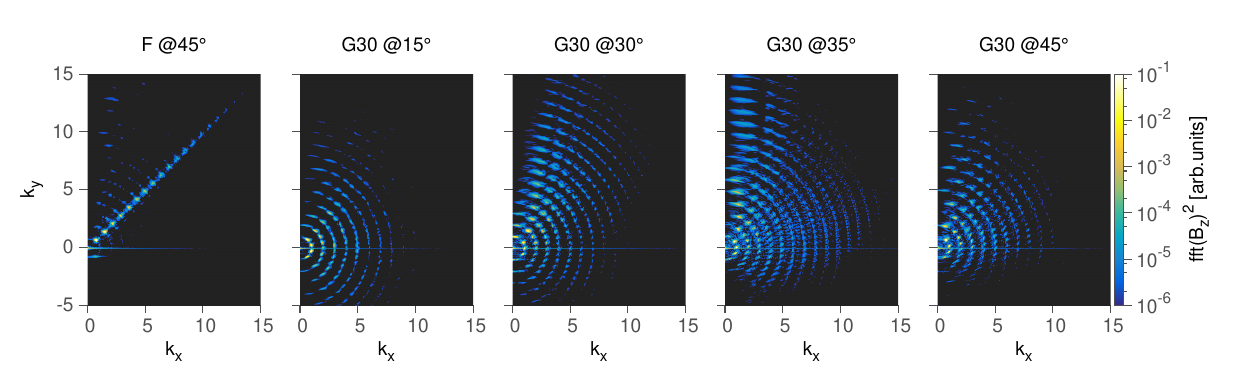}
\caption{Two-dimensional Fourier transform $\hat{B}_z(k_x,k_y)$ for F irradiated at $45^{\circ}$ and for G30 irradiated at $\theta_i=15^{\circ}, 30^{\circ}, 35^{\circ}, 45^{\circ}$. For F all the harmonics are emitted along the specular reflection direction. For G30 higher-order harmonics are emitted when the target is irradiated near resonance. The harmonic emission is particularly strong close to the target tangent for $\theta_i= 35^{\circ}$. }
\label{fig:G3035zoom}
\end{figure}
\end{center}

\twocolumngrid
Fig.\ref{fig:bz} shows two snapshots of the magnetic field $B_z$ (perpendicular to the simulation plane) which is the most convenient choice to represent the spatial distribution of the EM fields. The snapshots are shown at the initial time $t =0 T$, when the laser pulse has not reached the target yet and $t=35 T$ (where $T=\lambda/c$) when the interaction with the target is over. The leftmost border of the target is at $x=0$. The $n=-1$ and $n=-2$ diffraction is apparent in Fig.\ref{fig:bz}. 
When a grating target is irradiated with an incidence angle $\theta_{\mbox{\tiny i}}$ close to the expected value for $\theta_{\mbox{\tiny res}}$, an enhancement of the local field intensity with respect to flat targets is observed. Fig.\ref{fig:largefig} shows that for an angle of incidence $\approx 35^{\circ}$ the maximum field is up to $\sim 5$ times the laser field, i.e. a factor 2.5 higher than with flat targets.
The degree of field enhancement which is observed also at angles quite different from $\theta_{\mbox{\tiny res}}$, i.e. far from the expected resonance, may be ascribed to the modulation of the field produced by reflection from a sinusoidal grating.

The angular spectrum of the emitted HHs was analyzed as follows. A 2D spatial Fourier transform of $B_z(x<0,y,t=35T)$, i.e. of the $B_z$ field in the vacuum region and after the interaction, is performed. 
The resulting Fourier transform $\hat{B_z}(k_x,k_y)$ is shown in Fig.\ref{fig:largefig}b) for the grating target irradiated at $35^{\circ}$. The dark spots on the $k=1$ ring correspond to the diffraction orders of the laser beam. The HH wavevectors appear as discrete spots on concentric rings corresponding to various harmonic orders. Most of the wavevectors are in the directions predicted by Eq.(\ref{eq:diffraction}) (which are marked with red points}). 
\begin{figure}
\includegraphics[width=1.0\columnwidth]{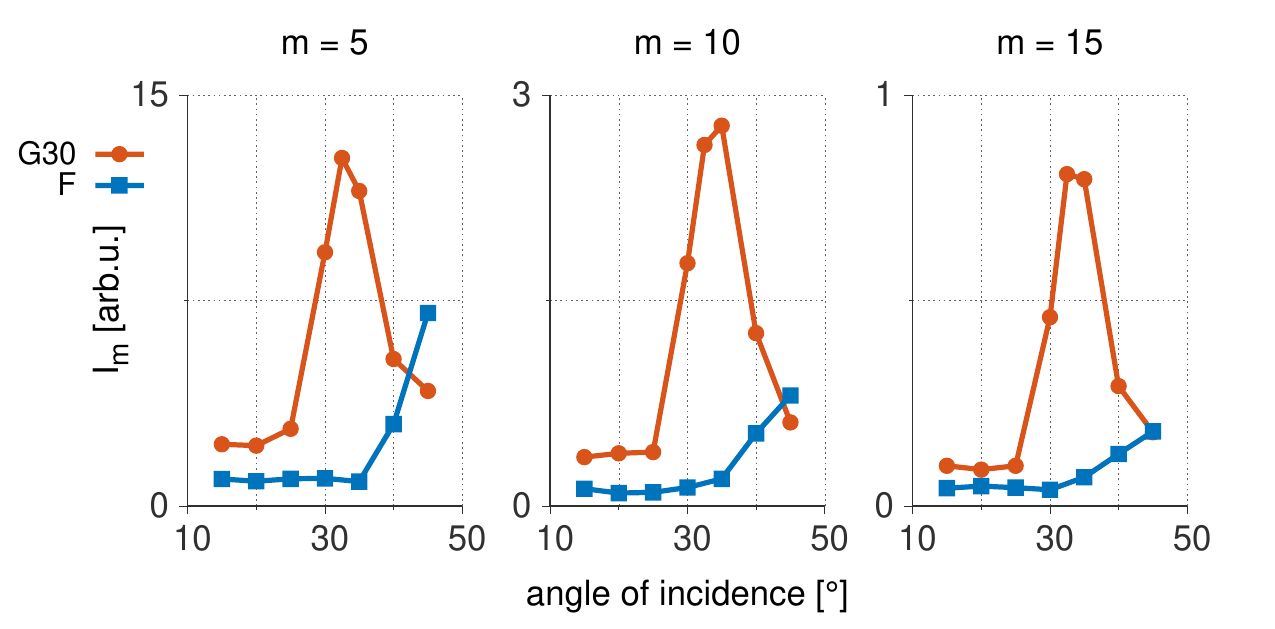}
\caption{
The total intensity $I_{m}$ (integrated over the emission angle) of the harmonics of order $m=5,10,15$ as a function of the angle of incidence for both F and G30 targets. In all three cases $I_m$ peaks at $35^{\circ}$ for G30, while for F $I_m$ increases monotonically with the incidence angle.}
\label{fig:maxharm}
\end{figure}
Fig.\ref{fig:G3035zoom} shows the same kind of graph for several simulations: G30 irradiated at $15^{\circ}, 30^{\circ}$ (the expected resonance angle) $, 35^{\circ}, 45^{\circ}$ and F irradiated at $45^{\circ}$.
For the F targets HHs are emitted almost exclusively along the direction of specular reflection: along such direction the $m=1$ order (i.e. the reflected laser pulse) is much more intense than the subsequent harmonic orders, which may be a problem in order to exploit HHs for applications.  In contrast, for the G30 target the HHs are angularly dispersed in different directions depending on the harmonic order. For an angle of incidence of $35^{\circ}$ the HHs are particularly intense in a direction close to the tangent to the surface. Such intense HH emission is well separated from the diffracted light at the fundamental frequency.
The stripe at $k_y = 0$ which is visible in all the panels of Fig.\ref{fig:G3035zoom} can be attributed to the development of a quasi-static magnetic field localized at the target surface\cite{BigongiariPoP2011}.

The angular distribution of the $m$-th harmonic in the direction $\theta$ {is} calculated as follows, 
\begin{equation}
\frac{\mbox{d}I_m(\theta)}{\mbox{d}\theta} = \int_{k_m-1/2}^{k_m+ 1/2} k\mbox{d}k |\hat{B_z}(k_x,k_y)|^2 \, ,
\label{eq:dImdQ}
\end{equation}
where $k_x = k\cos(\theta)$ and $k_y = k\sin(\theta)$.  The total $m$-th harmonic yield $I_{m}=I_m(\theta)$ is given by the integration of Eq.(\ref{eq:dImdQ}) 
over the full angle {($|\theta|<{\pi/2}$)}.
Fig.\ref{fig:maxharm} shows a comparison of $I_{m}$ between F and G30 targets, for three different harmonic orders ($m=5,10,15$). A prominent maximum is observed for the HHs emitted for G30 at an angle slightly larger than the one expected for surface plasmon resonance according to the linear theory ($32.5^{\circ}-35^{\circ}$ rather than $30^{\circ}$), in agreement with previous experimental and numerical observations on SP-driven electron acceleration at very high laser intensities\cite{Fedeli_PRL2016}; thus, the discrepancy might be ascribed to nonlinear and relativistic effects. The overall enhancement of harmonic emission shows that the effect of SP excitation overlaps with the diffraction from the grating.
The value of $I_{m}$ at the peak for G30 exceeds by more than one order of magnitude the value for the F target at the same angle of incidence $\theta_{\mbox{\tiny i}}$. In contrast, at larger values of $\theta_{\mbox{\tiny i}}$ the harmonic yield for the F target becomes larger than for G30. This might be explained by the ``shadow'' effect of the grating at large angles of incidence, well out of resonance. 

\begin{figure}
\centering
\includegraphics[width=\columnwidth]{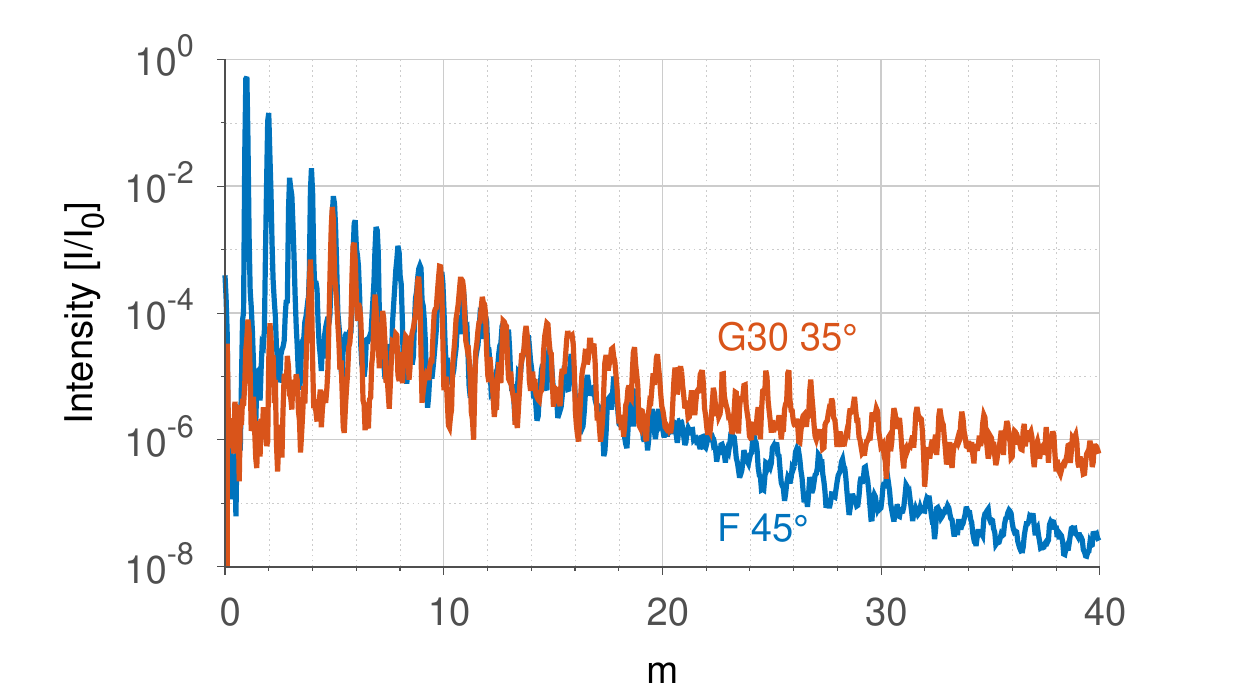}
\caption{Harmonic spectrum collected at $45 \pm 2.5 ^{\circ}$ for F irradiated at $45^{\circ}$ and at $80 \pm 2.5 ^\circ$ for G30 irradiated at $35^{\circ}$. The intensity is normalized with respect to that of the incoming beam.
}
\label{fig:compig}
\end{figure}

The SP-driven enhancement of HHs appears to be stronger with increasing harmonic order $m$. To investigate this behavior further, two cases (G30 irradiated at $35^{\circ}$ and F irradiated at $45^{\circ}$) were simulated with an increased resolution of $\lambda/400$  to compute HH generation reliably up to $m \approx 40$.  These cases were selected because they provide the highest harmonic yield for the two target types. Harmonic spectra collected at $45^{\circ}$ for F and at $80^{\circ}$ for G30 are shown in Fig.\ref{fig:compig}. As expected the spectrum for F is dominated by the $m=1$ harmonic (whose intensity is $\approx 60 \% $ of that of the incoming beam), while the $m=2$ harmonic is $\approx 6$ times less intense.
On the contrary, for G30 we observe a significant increase of the intensity of emitted harmonics, which is particularly evident for higher orders, being $\approx$ 2 orders of magnitude for $m = 40$.

In conclusion, the SP-driven enhancement of high harmonic emission from grating targets irradiated at relativistic intensities has been studied by numerical simulations.
When gratings are irradiated with an angle of incidence  close to the one expected for surface plasmon resonance, the intensity emitted in a given harmonic order increases significantly, the effect being stronger with increasing harmonic order.
In particular, a strong emission close to the target tangent is observed.  
This suggests that in an actual experiment it should be possible to detect more harmonic orders with the grating target irradiated near resonance than with a flat target.
Implementation of the present scheme may allow the generation of intense, spatially separated high harmonics. 
High repetition rate operation may be achieved by using engraved tape targets with a suitable choice of the target material \cite{shawJAP13,BorotOL11}, or even developing all-optical generation of plasma gratings \cite{monchocePRL14}.
The study also gives a further demonstration of the exploitation of plasmonic effects in the high field regime to manipulate laser-matter interactions.
\begin{acknowledgments}
We acknowledge ISCRA for the access to the Intel cluster Galileo, based in Italy at CINECA, via the project ``LaCoSA''. We gratefully acknowledge computational support from HPC Cluster CNAF (Bologna, Italy) and we wish to thank the HPC team for their assistance. The author L. Fedeli has been partially supported by European Research Council Consolidator Grant ENSURE (ERC-2014-CoG No. 647554) 
\end{acknowledgments}


\begin{thebibliography}{55}%
\makeatletter
\providecommand \@ifxundefined [1]{%
 \@ifx{#1\undefined}
}%
\providecommand \@ifnum [1]{%
 \ifnum #1\expandafter \@firstoftwo
 \else \expandafter \@secondoftwo
 \fi
}%
\providecommand \@ifx [1]{%
 \ifx #1\expandafter \@firstoftwo
 \else \expandafter \@secondoftwo
 \fi
}%
\providecommand \natexlab [1]{#1}%
\providecommand \enquote  [1]{``#1''}%
\providecommand \bibnamefont  [1]{#1}%
\providecommand \bibfnamefont [1]{#1}%
\providecommand \citenamefont [1]{#1}%
\providecommand \href@noop [0]{\@secondoftwo}%
\providecommand \href [0]{\begingroup \@sanitize@url \@href}%
\providecommand \@href[1]{\@@startlink{#1}\@@href}%
\providecommand \@@href[1]{\endgroup#1\@@endlink}%
\providecommand \@sanitize@url [0]{\catcode `\\12\catcode `\$12\catcode
  `\&12\catcode `\#12\catcode `\^12\catcode `\_12\catcode `\%12\relax}%
\providecommand \@@startlink[1]{}%
\providecommand \@@endlink[0]{}%
\providecommand \url  [0]{\begingroup\@sanitize@url \@url }%
\providecommand \@url [1]{\endgroup\@href {#1}{\urlprefix }}%
\providecommand \urlprefix  [0]{URL }%
\providecommand \Eprint [0]{\href }%
\providecommand \doibase [0]{http://dx.doi.org/}%
\providecommand \selectlanguage [0]{\@gobble}%
\providecommand \bibinfo  [0]{\@secondoftwo}%
\providecommand \bibfield  [0]{\@secondoftwo}%
\providecommand \translation [1]{[#1]}%
\providecommand \BibitemOpen [0]{}%
\providecommand \bibitemStop [0]{}%
\providecommand \bibitemNoStop [0]{.\EOS\space}%
\providecommand \EOS [0]{\spacefactor3000\relax}%
\providecommand \BibitemShut  [1]{\csname bibitem#1\endcsname}%
\let\auto@bib@innerbib\@empty
\bibitem [{\citenamefont {Sansone}, \citenamefont {Poletto},\ and\
  \citenamefont {Nisoli}(2011)}]{Sansone_NatPhot2011}%
  \BibitemOpen
  \bibfield  {author} {\bibinfo {author} {\bibfnamefont {G.}~\bibnamefont
  {Sansone}}, \bibinfo {author} {\bibfnamefont {L.}~\bibnamefont {Poletto}}, \
  and\ \bibinfo {author} {\bibfnamefont {M.}~\bibnamefont {Nisoli}},\
  }\href@noop {} {\bibfield  {journal} {\bibinfo  {journal} {Nature Photonics}\
  }\textbf {\bibinfo {volume} {5}},\ \bibinfo {pages} {655} (\bibinfo {year}
  {2011})}\BibitemShut {NoStop}%
\bibitem [{\citenamefont {Krausz}\ and\ \citenamefont
  {Ivanov}(2009)}]{Krausz_RevModPhys2009}%
  \BibitemOpen
  \bibfield  {author} {\bibinfo {author} {\bibfnamefont {F.}~\bibnamefont
  {Krausz}}\ and\ \bibinfo {author} {\bibfnamefont {M.}~\bibnamefont
  {Ivanov}},\ }\href@noop {} {\bibfield  {journal} {\bibinfo  {journal}
  {Reviews of Modern Physics}\ }\textbf {\bibinfo {volume} {81}},\ \bibinfo
  {pages} {163} (\bibinfo {year} {2009})}\BibitemShut {NoStop}%
\bibitem [{\citenamefont {Paul}\ \emph {et~al.}(2003)\citenamefont {Paul},
  \citenamefont {Bartels}, \citenamefont {Tobey}, \citenamefont {Green},
  \citenamefont {Weiman}, \citenamefont {Christov}, \citenamefont {Murnane},
  \citenamefont {Kapteyn},\ and\ \citenamefont {Backus}}]{paulN03}%
  \BibitemOpen
  \bibfield  {author} {\bibinfo {author} {\bibfnamefont {A.}~\bibnamefont
  {Paul}}, \bibinfo {author} {\bibfnamefont {R.~A.}\ \bibnamefont {Bartels}},
  \bibinfo {author} {\bibfnamefont {R.}~\bibnamefont {Tobey}}, \bibinfo
  {author} {\bibfnamefont {H.}~\bibnamefont {Green}}, \bibinfo {author}
  {\bibfnamefont {S.}~\bibnamefont {Weiman}}, \bibinfo {author} {\bibfnamefont
  {I.~P.}\ \bibnamefont {Christov}}, \bibinfo {author} {\bibfnamefont {M.~M.}\
  \bibnamefont {Murnane}}, \bibinfo {author} {\bibfnamefont {H.~C.}\
  \bibnamefont {Kapteyn}}, \ and\ \bibinfo {author} {\bibfnamefont
  {S.}~\bibnamefont {Backus}},\ }\href {\doibase 10.1038/nature01222}
  {\bibfield  {journal} {\bibinfo  {journal} {Nature}\ }\textbf {\bibinfo
  {volume} {421}},\ \bibinfo {pages} {51} (\bibinfo {year} {2003})}\BibitemShut
  {NoStop}%
\bibitem [{\citenamefont {Zepf}\ \emph {et~al.}(2007)\citenamefont {Zepf},
  \citenamefont {Dromey}, \citenamefont {Landreman}, \citenamefont {Foster},\
  and\ \citenamefont {Hooker}}]{zepfPRL07}%
  \BibitemOpen
  \bibfield  {author} {\bibinfo {author} {\bibfnamefont {M.}~\bibnamefont
  {Zepf}}, \bibinfo {author} {\bibfnamefont {B.}~\bibnamefont {Dromey}},
  \bibinfo {author} {\bibfnamefont {M.}~\bibnamefont {Landreman}}, \bibinfo
  {author} {\bibfnamefont {P.}~\bibnamefont {Foster}}, \ and\ \bibinfo {author}
  {\bibfnamefont {S.~M.}\ \bibnamefont {Hooker}},\ }\href {\doibase
  10.1103/PhysRevLett.99.143901} {\bibfield  {journal} {\bibinfo  {journal}
  {Phys. Rev. Lett.}\ }\textbf {\bibinfo {volume} {99}},\ \bibinfo {pages}
  {143901} (\bibinfo {year} {2007})}\BibitemShut {NoStop}%
\bibitem [{\citenamefont {Shiner}\ \emph {et~al.}(2009)\citenamefont {Shiner},
  \citenamefont {Trallero-Herrero}, \citenamefont {Kajumba}, \citenamefont
  {Bandulet}, \citenamefont {Comtois}, \citenamefont {L\'egar\'e},
  \citenamefont {Gigu\`ere}, \citenamefont {Kieffer}, \citenamefont {Corkum},\
  and\ \citenamefont {Villeneuve}}]{shinerPRL09}%
  \BibitemOpen
  \bibfield  {author} {\bibinfo {author} {\bibfnamefont {A.~D.}\ \bibnamefont
  {Shiner}}, \bibinfo {author} {\bibfnamefont {C.}~\bibnamefont
  {Trallero-Herrero}}, \bibinfo {author} {\bibfnamefont {N.}~\bibnamefont
  {Kajumba}}, \bibinfo {author} {\bibfnamefont {H.-C.}\ \bibnamefont
  {Bandulet}}, \bibinfo {author} {\bibfnamefont {D.}~\bibnamefont {Comtois}},
  \bibinfo {author} {\bibfnamefont {F.}~\bibnamefont {L\'egar\'e}}, \bibinfo
  {author} {\bibfnamefont {M.}~\bibnamefont {Gigu\`ere}}, \bibinfo {author}
  {\bibfnamefont {J.-C.}\ \bibnamefont {Kieffer}}, \bibinfo {author}
  {\bibfnamefont {P.~B.}\ \bibnamefont {Corkum}}, \ and\ \bibinfo {author}
  {\bibfnamefont {D.~M.}\ \bibnamefont {Villeneuve}},\ }\href {\doibase
  10.1103/PhysRevLett.103.073902} {\bibfield  {journal} {\bibinfo  {journal}
  {Phys. Rev. Lett.}\ }\textbf {\bibinfo {volume} {103}},\ \bibinfo {pages}
  {073902} (\bibinfo {year} {2009})}\BibitemShut {NoStop}%
\bibitem [{\citenamefont {Popmintchev}\ \emph {et~al.}(2015)\citenamefont
  {Popmintchev}, \citenamefont {Hern{\'a}ndez-Garc{\'\i}a}, \citenamefont
  {Dollar}, \citenamefont {Mancuso}, \citenamefont {P{\'e}rez-Hern{\'a}ndez},
  \citenamefont {Chen}, \citenamefont {Hankla}, \citenamefont {Gao},
  \citenamefont {Shim}, \citenamefont {Gaeta}, \citenamefont {Tarazkar},
  \citenamefont {Romanov}, \citenamefont {Levis}, \citenamefont {Gaffney},
  \citenamefont {Foord}, \citenamefont {Libby}, \citenamefont {Jaron-Becker},
  \citenamefont {Becker}, \citenamefont {Plaja}, \citenamefont {Murnane},
  \citenamefont {Kapteyn},\ and\ \citenamefont {Popmintchev}}]{PopmintchevS15}%
  \BibitemOpen
  \bibfield  {author} {\bibinfo {author} {\bibfnamefont {D.}~\bibnamefont
  {Popmintchev}}, \bibinfo {author} {\bibfnamefont {C.}~\bibnamefont
  {Hern{\'a}ndez-Garc{\'\i}a}}, \bibinfo {author} {\bibfnamefont
  {F.}~\bibnamefont {Dollar}}, \bibinfo {author} {\bibfnamefont
  {C.}~\bibnamefont {Mancuso}}, \bibinfo {author} {\bibfnamefont {J.~A.}\
  \bibnamefont {P{\'e}rez-Hern{\'a}ndez}}, \bibinfo {author} {\bibfnamefont
  {M.-C.}\ \bibnamefont {Chen}}, \bibinfo {author} {\bibfnamefont
  {A.}~\bibnamefont {Hankla}}, \bibinfo {author} {\bibfnamefont
  {X.}~\bibnamefont {Gao}}, \bibinfo {author} {\bibfnamefont {B.}~\bibnamefont
  {Shim}}, \bibinfo {author} {\bibfnamefont {A.~L.}\ \bibnamefont {Gaeta}},
  \bibinfo {author} {\bibfnamefont {M.}~\bibnamefont {Tarazkar}}, \bibinfo
  {author} {\bibfnamefont {D.~A.}\ \bibnamefont {Romanov}}, \bibinfo {author}
  {\bibfnamefont {R.~J.}\ \bibnamefont {Levis}}, \bibinfo {author}
  {\bibfnamefont {J.~A.}\ \bibnamefont {Gaffney}}, \bibinfo {author}
  {\bibfnamefont {M.}~\bibnamefont {Foord}}, \bibinfo {author} {\bibfnamefont
  {S.~B.}\ \bibnamefont {Libby}}, \bibinfo {author} {\bibfnamefont
  {A.}~\bibnamefont {Jaron-Becker}}, \bibinfo {author} {\bibfnamefont
  {A.}~\bibnamefont {Becker}}, \bibinfo {author} {\bibfnamefont
  {L.}~\bibnamefont {Plaja}}, \bibinfo {author} {\bibfnamefont {M.~M.}\
  \bibnamefont {Murnane}}, \bibinfo {author} {\bibfnamefont {H.~C.}\
  \bibnamefont {Kapteyn}}, \ and\ \bibinfo {author} {\bibfnamefont
  {T.}~\bibnamefont {Popmintchev}},\ }\href {\doibase 10.1126/science.aac9755}
  {\bibfield  {journal} {\bibinfo  {journal} {Science}\ }\textbf {\bibinfo
  {volume} {350}},\ \bibinfo {pages} {1225} (\bibinfo {year}
  {2015})}\BibitemShut {NoStop}%
\bibitem [{\citenamefont {von~der Linde}\ \emph {et~al.}(1995)\citenamefont
  {von~der Linde}, \citenamefont {Engers}, \citenamefont {Jenke}, \citenamefont
  {Agostini}, \citenamefont {Grillon}, \citenamefont {Nibbering}, \citenamefont
  {Mysyrowicz},\ and\ \citenamefont {Antonetti}}]{VonDerLinde_PRA1995}%
  \BibitemOpen
  \bibfield  {author} {\bibinfo {author} {\bibfnamefont {D.}~\bibnamefont
  {von~der Linde}}, \bibinfo {author} {\bibfnamefont {T.}~\bibnamefont
  {Engers}}, \bibinfo {author} {\bibfnamefont {G.}~\bibnamefont {Jenke}},
  \bibinfo {author} {\bibfnamefont {P.}~\bibnamefont {Agostini}}, \bibinfo
  {author} {\bibfnamefont {G.}~\bibnamefont {Grillon}}, \bibinfo {author}
  {\bibfnamefont {E.}~\bibnamefont {Nibbering}}, \bibinfo {author}
  {\bibfnamefont {A.}~\bibnamefont {Mysyrowicz}}, \ and\ \bibinfo {author}
  {\bibfnamefont {A.}~\bibnamefont {Antonetti}},\ }\href {\doibase
  10.1103/PhysRevA.52.R25} {\bibfield  {journal} {\bibinfo  {journal} {Phys.
  Rev. A}\ }\textbf {\bibinfo {volume} {52}},\ \bibinfo {pages} {R25} (\bibinfo
  {year} {1995})}\BibitemShut {NoStop}%
\bibitem [{\citenamefont {Norreys}\ \emph {et~al.}(1996)\citenamefont
  {Norreys}, \citenamefont {Zepf}, \citenamefont {Moustaizis}, \citenamefont
  {Fews}, \citenamefont {Zhang}, \citenamefont {Lee}, \citenamefont
  {Bakarezos}, \citenamefont {Danson}, \citenamefont {Dyson}, \citenamefont
  {Gibbon}, \citenamefont {Loukakos}, \citenamefont {Neely}, \citenamefont
  {Walsh}, \citenamefont {Wark},\ and\ \citenamefont
  {Dangor}}]{Norreys_PRL1996}%
  \BibitemOpen
  \bibfield  {author} {\bibinfo {author} {\bibfnamefont {P.~A.}\ \bibnamefont
  {Norreys}}, \bibinfo {author} {\bibfnamefont {M.}~\bibnamefont {Zepf}},
  \bibinfo {author} {\bibfnamefont {S.}~\bibnamefont {Moustaizis}}, \bibinfo
  {author} {\bibfnamefont {A.~P.}\ \bibnamefont {Fews}}, \bibinfo {author}
  {\bibfnamefont {J.}~\bibnamefont {Zhang}}, \bibinfo {author} {\bibfnamefont
  {P.}~\bibnamefont {Lee}}, \bibinfo {author} {\bibfnamefont {M.}~\bibnamefont
  {Bakarezos}}, \bibinfo {author} {\bibfnamefont {C.~N.}\ \bibnamefont
  {Danson}}, \bibinfo {author} {\bibfnamefont {A.}~\bibnamefont {Dyson}},
  \bibinfo {author} {\bibfnamefont {P.}~\bibnamefont {Gibbon}}, \bibinfo
  {author} {\bibfnamefont {P.}~\bibnamefont {Loukakos}}, \bibinfo {author}
  {\bibfnamefont {D.}~\bibnamefont {Neely}}, \bibinfo {author} {\bibfnamefont
  {F.~N.}\ \bibnamefont {Walsh}}, \bibinfo {author} {\bibfnamefont {J.~S.}\
  \bibnamefont {Wark}}, \ and\ \bibinfo {author} {\bibfnamefont {A.~E.}\
  \bibnamefont {Dangor}},\ }\href {\doibase 10.1103/PhysRevLett.76.1832}
  {\bibfield  {journal} {\bibinfo  {journal} {Phys. Rev. Lett.}\ }\textbf
  {\bibinfo {volume} {76}},\ \bibinfo {pages} {1832} (\bibinfo {year}
  {1996})}\BibitemShut {NoStop}%
\bibitem [{\citenamefont {Tarasevitch}\ \emph {et~al.}(2000)\citenamefont
  {Tarasevitch}, \citenamefont {Orisch}, \citenamefont {von~der Linde},
  \citenamefont {Balcou}, \citenamefont {Rey}, \citenamefont {Chambaret},
  \citenamefont {Teubner}, \citenamefont {Kl\"opfel},\ and\ \citenamefont
  {Theobald}}]{Tarasevitch_PRA2000}%
  \BibitemOpen
  \bibfield  {author} {\bibinfo {author} {\bibfnamefont {A.}~\bibnamefont
  {Tarasevitch}}, \bibinfo {author} {\bibfnamefont {A.}~\bibnamefont {Orisch}},
  \bibinfo {author} {\bibfnamefont {D.}~\bibnamefont {von~der Linde}}, \bibinfo
  {author} {\bibfnamefont {P.}~\bibnamefont {Balcou}}, \bibinfo {author}
  {\bibfnamefont {G.}~\bibnamefont {Rey}}, \bibinfo {author} {\bibfnamefont
  {J.-P.}\ \bibnamefont {Chambaret}}, \bibinfo {author} {\bibfnamefont
  {U.}~\bibnamefont {Teubner}}, \bibinfo {author} {\bibfnamefont
  {D.}~\bibnamefont {Kl\"opfel}}, \ and\ \bibinfo {author} {\bibfnamefont
  {W.}~\bibnamefont {Theobald}},\ }\href {\doibase 10.1103/PhysRevA.62.023816}
  {\bibfield  {journal} {\bibinfo  {journal} {Phys. Rev. A}\ }\textbf {\bibinfo
  {volume} {62}},\ \bibinfo {pages} {023816} (\bibinfo {year}
  {2000})}\BibitemShut {NoStop}%
\bibitem [{\citenamefont {Teubner}\ and\ \citenamefont
  {Gibbon}(2009)}]{Teubner_RevModPhys2009}%
  \BibitemOpen
  \bibfield  {author} {\bibinfo {author} {\bibfnamefont {U.}~\bibnamefont
  {Teubner}}\ and\ \bibinfo {author} {\bibfnamefont {P.}~\bibnamefont
  {Gibbon}},\ }\href {\doibase 10.1103/RevModPhys.81.445} {\bibfield  {journal}
  {\bibinfo  {journal} {Rev. Mod. Phys.}\ }\textbf {\bibinfo {volume} {81}},\
  \bibinfo {pages} {445} (\bibinfo {year} {2009})}\BibitemShut {NoStop}%
\bibitem [{\citenamefont {Thaury}\ and\ \citenamefont
  {Qu\'{e}r\'{e}}(2010)}]{Thaury_JPB2010}%
  \BibitemOpen
  \bibfield  {author} {\bibinfo {author} {\bibfnamefont {C.}~\bibnamefont
  {Thaury}}\ and\ \bibinfo {author} {\bibfnamefont {F.}~\bibnamefont
  {Qu\'{e}r\'{e}}},\ }\href {http://stacks.iop.org/0953-4075/43/i=21/a=213001}
  {\bibfield  {journal} {\bibinfo  {journal} {Journal of Physics B: Atomic,
  Molecular and Optical Physics}\ }\textbf {\bibinfo {volume} {43}},\ \bibinfo
  {pages} {213001} (\bibinfo {year} {2010})}\BibitemShut {NoStop}%
\bibitem [{\citenamefont {Bierbach}\ \emph {et~al.}(2012)\citenamefont
  {Bierbach}, \citenamefont {R{\"o}del}, \citenamefont {Yeung}, \citenamefont
  {Dromey}, \citenamefont {Hahn}, \citenamefont {Pour}, \citenamefont {Fuchs},
  \citenamefont {Paz}, \citenamefont {Herzer}, \citenamefont {Kuschel},
  \citenamefont {J{\"a}ckel}, \citenamefont {Kaluza}, \citenamefont {Pretzler},
  \citenamefont {Zepf},\ and\ \citenamefont {Paulus}}]{Bierbach_NJP2012}%
  \BibitemOpen
  \bibfield  {author} {\bibinfo {author} {\bibfnamefont {J.}~\bibnamefont
  {Bierbach}}, \bibinfo {author} {\bibfnamefont {C.}~\bibnamefont {R{\"o}del}},
  \bibinfo {author} {\bibfnamefont {M.}~\bibnamefont {Yeung}}, \bibinfo
  {author} {\bibfnamefont {B.}~\bibnamefont {Dromey}}, \bibinfo {author}
  {\bibfnamefont {T.}~\bibnamefont {Hahn}}, \bibinfo {author} {\bibfnamefont
  {A.~G.}\ \bibnamefont {Pour}}, \bibinfo {author} {\bibfnamefont
  {S.}~\bibnamefont {Fuchs}}, \bibinfo {author} {\bibfnamefont {A.~E.}\
  \bibnamefont {Paz}}, \bibinfo {author} {\bibfnamefont {S.}~\bibnamefont
  {Herzer}}, \bibinfo {author} {\bibfnamefont {S.}~\bibnamefont {Kuschel}},
  \bibinfo {author} {\bibfnamefont {O.}~\bibnamefont {J{\"a}ckel}}, \bibinfo
  {author} {\bibfnamefont {M.~C.}\ \bibnamefont {Kaluza}}, \bibinfo {author}
  {\bibfnamefont {G.}~\bibnamefont {Pretzler}}, \bibinfo {author}
  {\bibfnamefont {M.}~\bibnamefont {Zepf}}, \ and\ \bibinfo {author}
  {\bibfnamefont {G.~G.}\ \bibnamefont {Paulus}},\ }\href
  {http://stacks.iop.org/1367-2630/14/i=6/a=065005} {\bibfield  {journal}
  {\bibinfo  {journal} {New Journal of Physics}\ }\textbf {\bibinfo {volume}
  {14}},\ \bibinfo {pages} {065005} (\bibinfo {year} {2012})}\BibitemShut
  {NoStop}%
\bibitem [{\citenamefont {an~der Br\"ugge}\ \emph {et~al.}(2012)\citenamefont
  {an~der Br\"ugge}, \citenamefont {Kumar}, \citenamefont {Pukhov},\ and\
  \citenamefont {R\"odel}}]{Brugger_PRL2012}%
  \BibitemOpen
  \bibfield  {author} {\bibinfo {author} {\bibfnamefont {D.}~\bibnamefont
  {an~der Br\"ugge}}, \bibinfo {author} {\bibfnamefont {N.}~\bibnamefont
  {Kumar}}, \bibinfo {author} {\bibfnamefont {A.}~\bibnamefont {Pukhov}}, \
  and\ \bibinfo {author} {\bibfnamefont {C.}~\bibnamefont {R\"odel}},\ }\href
  {\doibase 10.1103/PhysRevLett.108.125002} {\bibfield  {journal} {\bibinfo
  {journal} {Phys. Rev. Lett.}\ }\textbf {\bibinfo {volume} {108}},\ \bibinfo
  {pages} {125002} (\bibinfo {year} {2012})}\BibitemShut {NoStop}%
\bibitem [{\citenamefont {Kahaly}\ \emph {et~al.}(2013)\citenamefont {Kahaly},
  \citenamefont {Monchoc\'e}, \citenamefont {Vincenti}, \citenamefont
  {Dzelzainis}, \citenamefont {Dromey}, \citenamefont {Zepf}, \citenamefont
  {Martin},\ and\ \citenamefont {Qu\'er\'e}}]{Kahaly_PRL2013}%
  \BibitemOpen
  \bibfield  {author} {\bibinfo {author} {\bibfnamefont {S.}~\bibnamefont
  {Kahaly}}, \bibinfo {author} {\bibfnamefont {S.}~\bibnamefont {Monchoc\'e}},
  \bibinfo {author} {\bibfnamefont {H.}~\bibnamefont {Vincenti}}, \bibinfo
  {author} {\bibfnamefont {T.}~\bibnamefont {Dzelzainis}}, \bibinfo {author}
  {\bibfnamefont {B.}~\bibnamefont {Dromey}}, \bibinfo {author} {\bibfnamefont
  {M.}~\bibnamefont {Zepf}}, \bibinfo {author} {\bibfnamefont {P.}~\bibnamefont
  {Martin}}, \ and\ \bibinfo {author} {\bibfnamefont {F.}~\bibnamefont
  {Qu\'er\'e}},\ }\href {\doibase 10.1103/PhysRevLett.110.175001} {\bibfield
  {journal} {\bibinfo  {journal} {Phys. Rev. Lett.}\ }\textbf {\bibinfo
  {volume} {110}},\ \bibinfo {pages} {175001} (\bibinfo {year}
  {2013})}\BibitemShut {NoStop}%
\bibitem [{\citenamefont {Heissler}\ \emph {et~al.}(2014)\citenamefont
  {Heissler}, \citenamefont {Barna}, \citenamefont {Mikhailova}, \citenamefont
  {Ma}, \citenamefont {Khrennikov}, \citenamefont {Karsch}, \citenamefont
  {Veisz}, \citenamefont {F{\"o}ldes},\ and\ \citenamefont
  {Tsakiris}}]{Heissler_APB2014}%
  \BibitemOpen
  \bibfield  {author} {\bibinfo {author} {\bibfnamefont {P.}~\bibnamefont
  {Heissler}}, \bibinfo {author} {\bibfnamefont {A.}~\bibnamefont {Barna}},
  \bibinfo {author} {\bibfnamefont {J.~M.}\ \bibnamefont {Mikhailova}},
  \bibinfo {author} {\bibfnamefont {G.}~\bibnamefont {Ma}}, \bibinfo {author}
  {\bibfnamefont {K.}~\bibnamefont {Khrennikov}}, \bibinfo {author}
  {\bibfnamefont {S.}~\bibnamefont {Karsch}}, \bibinfo {author} {\bibfnamefont
  {L.}~\bibnamefont {Veisz}}, \bibinfo {author} {\bibfnamefont {I.~B.}\
  \bibnamefont {F{\"o}ldes}}, \ and\ \bibinfo {author} {\bibfnamefont {G.~D.}\
  \bibnamefont {Tsakiris}},\ }\href {\doibase 10.1007/s00340-014-5968-x}
  {\bibfield  {journal} {\bibinfo  {journal} {Applied Physics B}\ }\textbf
  {\bibinfo {volume} {118}},\ \bibinfo {pages} {195} (\bibinfo {year}
  {2014})}\BibitemShut {NoStop}%
\bibitem [{\citenamefont {Bocoum}\ \emph {et~al.}(2016)\citenamefont {Bocoum},
  \citenamefont {Th\'evenet}, \citenamefont {B\"ohle}, \citenamefont
  {Beaurepaire}, \citenamefont {Vernier}, \citenamefont {Jullien},
  \citenamefont {Faure},\ and\ \citenamefont
  {Lopez-Martens}}]{PhysRevLett.116.185001}%
  \BibitemOpen
  \bibfield  {author} {\bibinfo {author} {\bibfnamefont {M.}~\bibnamefont
  {Bocoum}}, \bibinfo {author} {\bibfnamefont {M.}~\bibnamefont {Th\'evenet}},
  \bibinfo {author} {\bibfnamefont {F.}~\bibnamefont {B\"ohle}}, \bibinfo
  {author} {\bibfnamefont {B.}~\bibnamefont {Beaurepaire}}, \bibinfo {author}
  {\bibfnamefont {A.}~\bibnamefont {Vernier}}, \bibinfo {author} {\bibfnamefont
  {A.}~\bibnamefont {Jullien}}, \bibinfo {author} {\bibfnamefont
  {J.}~\bibnamefont {Faure}}, \ and\ \bibinfo {author} {\bibfnamefont
  {R.}~\bibnamefont {Lopez-Martens}},\ }\href {\doibase
  10.1103/PhysRevLett.116.185001} {\bibfield  {journal} {\bibinfo  {journal}
  {Phys. Rev. Lett.}\ }\textbf {\bibinfo {volume} {116}},\ \bibinfo {pages}
  {185001} (\bibinfo {year} {2016})}\BibitemShut {NoStop}%
\bibitem [{\citenamefont {Lichters}, \citenamefont {{Meyer-ter-Vehn}},\ and\
  \citenamefont {Pukhov}(1996)}]{Lichters_PoP1996}%
  \BibitemOpen
  \bibfield  {author} {\bibinfo {author} {\bibfnamefont {R.}~\bibnamefont
  {Lichters}}, \bibinfo {author} {\bibfnamefont {J.}~\bibnamefont
  {{Meyer-ter-Vehn}}}, \ and\ \bibinfo {author} {\bibfnamefont
  {A.}~\bibnamefont {Pukhov}},\ }\href {\doibase
  http://dx.doi.org/10.1063/1.871619} {\bibfield  {journal} {\bibinfo
  {journal} {Phys. Plasmas}\ }\textbf {\bibinfo {volume} {3}},\ \bibinfo
  {pages} {3425} (\bibinfo {year} {1996})}\BibitemShut {NoStop}%
\bibitem [{\citenamefont {Baeva}, \citenamefont {Gordienko},\ and\
  \citenamefont {Pukhov}(2006)}]{Baeva_PRE2006}%
  \BibitemOpen
  \bibfield  {author} {\bibinfo {author} {\bibfnamefont {T.}~\bibnamefont
  {Baeva}}, \bibinfo {author} {\bibfnamefont {S.}~\bibnamefont {Gordienko}}, \
  and\ \bibinfo {author} {\bibfnamefont {A.}~\bibnamefont {Pukhov}},\ }\href
  {\doibase 10.1103/PhysRevE.74.046404} {\bibfield  {journal} {\bibinfo
  {journal} {Phys. Rev. E}\ }\textbf {\bibinfo {volume} {74}},\ \bibinfo
  {pages} {046404} (\bibinfo {year} {2006})}\BibitemShut {NoStop}%
\bibitem [{\citenamefont {Qu\'er\'e}\ \emph {et~al.}(2006)\citenamefont
  {Qu\'er\'e}, \citenamefont {Thaury}, \citenamefont {Monot}, \citenamefont
  {Dobosz}, \citenamefont {Martin}, \citenamefont {Geindre},\ and\
  \citenamefont {Audebert}}]{QuereHarm}%
  \BibitemOpen
  \bibfield  {author} {\bibinfo {author} {\bibfnamefont {F.}~\bibnamefont
  {Qu\'er\'e}}, \bibinfo {author} {\bibfnamefont {C.}~\bibnamefont {Thaury}},
  \bibinfo {author} {\bibfnamefont {P.}~\bibnamefont {Monot}}, \bibinfo
  {author} {\bibfnamefont {S.}~\bibnamefont {Dobosz}}, \bibinfo {author}
  {\bibfnamefont {P.}~\bibnamefont {Martin}}, \bibinfo {author} {\bibfnamefont
  {J.-P.}\ \bibnamefont {Geindre}}, \ and\ \bibinfo {author} {\bibfnamefont
  {P.}~\bibnamefont {Audebert}},\ }\href
  {http://link.aps.org/doi/10.1103/PhysRevLett.96.125004} {\bibfield  {journal}
  {\bibinfo  {journal} {Phys. Rev. Lett.}\ }\textbf {\bibinfo {volume} {96}},\
  \bibinfo {pages} {125004} (\bibinfo {year} {2006})}\BibitemShut {NoStop}%
\bibitem [{\citenamefont {Shaw}\ \emph {et~al.}(2013)\citenamefont {Shaw},
  \citenamefont {van Tilborg}, \citenamefont {Sokollik}, \citenamefont
  {Schroeder}, \citenamefont {McKinney}, \citenamefont {Artemiev},
  \citenamefont {Yashchuk}, \citenamefont {Gullikson},\ and\ \citenamefont
  {Leemans}}]{shawJAP13}%
  \BibitemOpen
  \bibfield  {author} {\bibinfo {author} {\bibfnamefont {B.~H.}\ \bibnamefont
  {Shaw}}, \bibinfo {author} {\bibfnamefont {J.}~\bibnamefont {van Tilborg}},
  \bibinfo {author} {\bibfnamefont {T.}~\bibnamefont {Sokollik}}, \bibinfo
  {author} {\bibfnamefont {C.~B.}\ \bibnamefont {Schroeder}}, \bibinfo {author}
  {\bibfnamefont {W.~R.}\ \bibnamefont {McKinney}}, \bibinfo {author}
  {\bibfnamefont {N.~A.}\ \bibnamefont {Artemiev}}, \bibinfo {author}
  {\bibfnamefont {V.~V.}\ \bibnamefont {Yashchuk}}, \bibinfo {author}
  {\bibfnamefont {E.~M.}\ \bibnamefont {Gullikson}}, \ and\ \bibinfo {author}
  {\bibfnamefont {W.~P.}\ \bibnamefont {Leemans}},\ }\href {\doibase
  10.1063/1.4816574} {\bibfield  {journal} {\bibinfo  {journal} {J. Appl.
  Phys.}\ }\textbf {\bibinfo {volume} {114}},\ \bibinfo {pages} {043106}
  (\bibinfo {year} {2013})}\BibitemShut {NoStop}%
\bibitem [{\citenamefont {Borot}\ \emph {et~al.}(2011)\citenamefont {Borot},
  \citenamefont {Malvache}, \citenamefont {Chen}, \citenamefont {Douillet},
  \citenamefont {Iaquianiello}, \citenamefont {Lefrou}, \citenamefont
  {Audebert}, \citenamefont {Geindre}, \citenamefont {Mourou}, \citenamefont
  {Qu\'{e}r\'{e}},\ and\ \citenamefont {Lopez-Martens}}]{BorotOL11}%
  \BibitemOpen
  \bibfield  {author} {\bibinfo {author} {\bibfnamefont {A.}~\bibnamefont
  {Borot}}, \bibinfo {author} {\bibfnamefont {A.}~\bibnamefont {Malvache}},
  \bibinfo {author} {\bibfnamefont {X.}~\bibnamefont {Chen}}, \bibinfo {author}
  {\bibfnamefont {D.}~\bibnamefont {Douillet}}, \bibinfo {author}
  {\bibfnamefont {G.}~\bibnamefont {Iaquianiello}}, \bibinfo {author}
  {\bibfnamefont {T.}~\bibnamefont {Lefrou}}, \bibinfo {author} {\bibfnamefont
  {P.}~\bibnamefont {Audebert}}, \bibinfo {author} {\bibfnamefont {J.-P.}\
  \bibnamefont {Geindre}}, \bibinfo {author} {\bibfnamefont {G.}~\bibnamefont
  {Mourou}}, \bibinfo {author} {\bibfnamefont {F.}~\bibnamefont
  {Qu\'{e}r\'{e}}}, \ and\ \bibinfo {author} {\bibfnamefont {R.}~\bibnamefont
  {Lopez-Martens}},\ }\href {\doibase 10.1364/OL.36.001461} {\bibfield
  {journal} {\bibinfo  {journal} {Opt. Lett.}\ }\textbf {\bibinfo {volume}
  {36}},\ \bibinfo {pages} {1461} (\bibinfo {year} {2011})}\BibitemShut
  {NoStop}%
\bibitem [{\citenamefont {Plaja}\ \emph {et~al.}(1998)\citenamefont {Plaja},
  \citenamefont {Roso}, \citenamefont {Rza?\.{z}ewski},\ and\ \citenamefont
  {Lewenstein}}]{Plaja_JOSAB_1998}%
  \BibitemOpen
  \bibfield  {author} {\bibinfo {author} {\bibfnamefont {L.}~\bibnamefont
  {Plaja}}, \bibinfo {author} {\bibfnamefont {L.}~\bibnamefont {Roso}},
  \bibinfo {author} {\bibfnamefont {K.}~\bibnamefont {Rza?\.{z}ewski}}, \ and\
  \bibinfo {author} {\bibfnamefont {M.}~\bibnamefont {Lewenstein}},\ }\href
  {\doibase 10.1364/JOSAB.15.001904} {\bibfield  {journal} {\bibinfo  {journal}
  {J. Opt. Soc. Am. B}\ }\textbf {\bibinfo {volume} {15}},\ \bibinfo {pages}
  {1904} (\bibinfo {year} {1998})}\BibitemShut {NoStop}%
\bibitem [{\citenamefont {Wheeler}\ \emph {et~al.}(2012)\citenamefont
  {Wheeler}, \citenamefont {Borot}, \citenamefont {Monchoc{\'e}}, \citenamefont
  {Vincenti}, \citenamefont {Ricci}, \citenamefont {Malvache}, \citenamefont
  {Lopez-Martens},\ and\ \citenamefont {Qu{\'e}r{\'e}}}]{Wheeler_NatPhot2012}%
  \BibitemOpen
  \bibfield  {author} {\bibinfo {author} {\bibfnamefont {J.~A.}\ \bibnamefont
  {Wheeler}}, \bibinfo {author} {\bibfnamefont {A.}~\bibnamefont {Borot}},
  \bibinfo {author} {\bibfnamefont {S.}~\bibnamefont {Monchoc{\'e}}}, \bibinfo
  {author} {\bibfnamefont {H.}~\bibnamefont {Vincenti}}, \bibinfo {author}
  {\bibfnamefont {A.}~\bibnamefont {Ricci}}, \bibinfo {author} {\bibfnamefont
  {A.}~\bibnamefont {Malvache}}, \bibinfo {author} {\bibfnamefont
  {R.}~\bibnamefont {Lopez-Martens}}, \ and\ \bibinfo {author} {\bibfnamefont
  {F.}~\bibnamefont {Qu{\'e}r{\'e}}},\ }\href@noop {} {\bibfield  {journal}
  {\bibinfo  {journal} {Nature Photonics}\ }\textbf {\bibinfo {volume} {6}},\
  \bibinfo {pages} {829} (\bibinfo {year} {2012})}\BibitemShut {NoStop}%
\bibitem [{\citenamefont {Vincenti}\ and\ \citenamefont
  {Qu\'er\'e}(2012)}]{Vincenti_PRL2012}%
  \BibitemOpen
  \bibfield  {author} {\bibinfo {author} {\bibfnamefont {H.}~\bibnamefont
  {Vincenti}}\ and\ \bibinfo {author} {\bibfnamefont {F.}~\bibnamefont
  {Qu\'er\'e}},\ }\href {\doibase 10.1103/PhysRevLett.108.113904} {\bibfield
  {journal} {\bibinfo  {journal} {Phys. Rev. Lett.}\ }\textbf {\bibinfo
  {volume} {108}},\ \bibinfo {pages} {113904} (\bibinfo {year}
  {2012})}\BibitemShut {NoStop}%
\bibitem [{\citenamefont {Yeung}\ \emph {et~al.}(2015)\citenamefont {Yeung},
  \citenamefont {Bierbach}, \citenamefont {Eckner}, \citenamefont {Rykovanov},
  \citenamefont {Kuschel}, \citenamefont {S\"avert}, \citenamefont
  {F{\"o}rster}, \citenamefont {R{\"o}del}, \citenamefont {Paulus},
  \citenamefont {Cousens}, \citenamefont {Coughlan}, \citenamefont {Dromey},\
  and\ \citenamefont {Zepf}}]{Yeung_PRL2015}%
  \BibitemOpen
  \bibfield  {author} {\bibinfo {author} {\bibfnamefont {M.}~\bibnamefont
  {Yeung}}, \bibinfo {author} {\bibfnamefont {J.}~\bibnamefont {Bierbach}},
  \bibinfo {author} {\bibfnamefont {E.}~\bibnamefont {Eckner}}, \bibinfo
  {author} {\bibfnamefont {S.}~\bibnamefont {Rykovanov}}, \bibinfo {author}
  {\bibfnamefont {S.}~\bibnamefont {Kuschel}}, \bibinfo {author} {\bibfnamefont
  {A.}~\bibnamefont {S\"avert}}, \bibinfo {author} {\bibfnamefont
  {M.}~\bibnamefont {F{\"o}rster}}, \bibinfo {author} {\bibfnamefont
  {C.}~\bibnamefont {R{\"o}del}}, \bibinfo {author} {\bibfnamefont {G.~G.}\
  \bibnamefont {Paulus}}, \bibinfo {author} {\bibfnamefont {S.}~\bibnamefont
  {Cousens}}, \bibinfo {author} {\bibfnamefont {M.}~\bibnamefont {Coughlan}},
  \bibinfo {author} {\bibfnamefont {B.}~\bibnamefont {Dromey}}, \ and\ \bibinfo
  {author} {\bibfnamefont {M.}~\bibnamefont {Zepf}},\ }\href {\doibase
  10.1103/PhysRevLett.115.193903} {\bibfield  {journal} {\bibinfo  {journal}
  {Phys. Rev. Lett.}\ }\textbf {\bibinfo {volume} {115}},\ \bibinfo {pages}
  {193903} (\bibinfo {year} {2015})}\BibitemShut {NoStop}%
\bibitem [{\citenamefont {Stamm}(2004)}]{0022-3727-37-23-005}%
  \BibitemOpen
  \bibfield  {author} {\bibinfo {author} {\bibfnamefont {U.}~\bibnamefont
  {Stamm}},\ }\href {http://stacks.iop.org/0022-3727/37/i=23/a=005} {\bibfield
  {journal} {\bibinfo  {journal} {Journal of Physics D: Applied Physics}\
  }\textbf {\bibinfo {volume} {37}},\ \bibinfo {pages} {3244} (\bibinfo {year}
  {2004})}\BibitemShut {NoStop}%
\bibitem [{\citenamefont {H{\"u}fner}(2013)}]{hufner2013photoelectron}%
  \BibitemOpen
  \bibfield  {author} {\bibinfo {author} {\bibfnamefont {S.}~\bibnamefont
  {H{\"u}fner}},\ }\href@noop {} {\emph {\bibinfo {title} {Photoelectron
  spectroscopy: principles and applications}}}\ (\bibinfo  {publisher}
  {Springer Science \& Business Media},\ \bibinfo {year} {2013})\BibitemShut
  {NoStop}%
\bibitem [{\citenamefont {Yeung}\ \emph {et~al.}(2011)\citenamefont {Yeung},
  \citenamefont {Zepf}, \citenamefont {Geissler},\ and\ \citenamefont
  {Dromey}}]{Yeung_OptLett2011}%
  \BibitemOpen
  \bibfield  {author} {\bibinfo {author} {\bibfnamefont {M.}~\bibnamefont
  {Yeung}}, \bibinfo {author} {\bibfnamefont {M.}~\bibnamefont {Zepf}},
  \bibinfo {author} {\bibfnamefont {M.}~\bibnamefont {Geissler}}, \ and\
  \bibinfo {author} {\bibfnamefont {B.}~\bibnamefont {Dromey}},\ }\href
  {\doibase 10.1364/OL.36.002333} {\bibfield  {journal} {\bibinfo  {journal}
  {Opt. Lett.}\ }\textbf {\bibinfo {volume} {36}},\ \bibinfo {pages} {2333}
  (\bibinfo {year} {2011})}\BibitemShut {NoStop}%
\bibitem [{\citenamefont {Yeung}\ \emph {et~al.}(2013)\citenamefont {Yeung},
  \citenamefont {Dromey}, \citenamefont {R{\"o}del}, \citenamefont {Bierbach},
  \citenamefont {W{\"u}nsche}, \citenamefont {Paulus}, \citenamefont {Hahn},
  \citenamefont {Hemmers}, \citenamefont {Stelzmann}, \citenamefont
  {Pretzler},\ and\ \citenamefont {Zepf}}]{Yeung_NJP2013}%
  \BibitemOpen
  \bibfield  {author} {\bibinfo {author} {\bibfnamefont {M.}~\bibnamefont
  {Yeung}}, \bibinfo {author} {\bibfnamefont {B.}~\bibnamefont {Dromey}},
  \bibinfo {author} {\bibfnamefont {C.}~\bibnamefont {R{\"o}del}}, \bibinfo
  {author} {\bibfnamefont {J.}~\bibnamefont {Bierbach}}, \bibinfo {author}
  {\bibfnamefont {M.}~\bibnamefont {W{\"u}nsche}}, \bibinfo {author}
  {\bibfnamefont {G.}~\bibnamefont {Paulus}}, \bibinfo {author} {\bibfnamefont
  {T.}~\bibnamefont {Hahn}}, \bibinfo {author} {\bibfnamefont {D.}~\bibnamefont
  {Hemmers}}, \bibinfo {author} {\bibfnamefont {C.}~\bibnamefont {Stelzmann}},
  \bibinfo {author} {\bibfnamefont {G.}~\bibnamefont {Pretzler}}, \ and\
  \bibinfo {author} {\bibfnamefont {M.}~\bibnamefont {Zepf}},\ }\href
  {http://stacks.iop.org/1367-2630/15/i=2/a=025042} {\bibfield  {journal}
  {\bibinfo  {journal} {New Journal of Physics}\ }\textbf {\bibinfo {volume}
  {15}},\ \bibinfo {pages} {025042} (\bibinfo {year} {2013})}\BibitemShut
  {NoStop}%
\bibitem [{\citenamefont {Pan}, \citenamefont {Zheng},\ and\ \citenamefont
  {He}(2016)}]{PanHarm2016}%
  \BibitemOpen
  \bibfield  {author} {\bibinfo {author} {\bibfnamefont {K.~Q.}\ \bibnamefont
  {Pan}}, \bibinfo {author} {\bibfnamefont {C.~Y.}\ \bibnamefont {Zheng}}, \
  and\ \bibinfo {author} {\bibfnamefont {X.~T.}\ \bibnamefont {He}},\
  }\href@noop {} {\bibfield  {journal} {\bibinfo  {journal} {Phys. Plasmas}\
  }\textbf {\bibinfo {volume} {23}},\ \bibinfo {eid} {023109} (\bibinfo {year}
  {2016})}\BibitemShut {NoStop}%
\bibitem [{\citenamefont {Zhang}\ \emph {et~al.}(2016)\citenamefont {Zhang},
  \citenamefont {Zhuo}, \citenamefont {Zou}, \citenamefont {Gan}, \citenamefont
  {Zhou}, \citenamefont {Li}, \citenamefont {Yu},\ and\ \citenamefont
  {Yu}}]{PhysRevE.93.053206}%
  \BibitemOpen
  \bibfield  {author} {\bibinfo {author} {\bibfnamefont {S.~J.}\ \bibnamefont
  {Zhang}}, \bibinfo {author} {\bibfnamefont {H.~B.}\ \bibnamefont {Zhuo}},
  \bibinfo {author} {\bibfnamefont {D.~B.}\ \bibnamefont {Zou}}, \bibinfo
  {author} {\bibfnamefont {L.~F.}\ \bibnamefont {Gan}}, \bibinfo {author}
  {\bibfnamefont {H.~Y.}\ \bibnamefont {Zhou}}, \bibinfo {author}
  {\bibfnamefont {X.~Z.}\ \bibnamefont {Li}}, \bibinfo {author} {\bibfnamefont
  {M.~Y.}\ \bibnamefont {Yu}}, \ and\ \bibinfo {author} {\bibfnamefont
  {W.}~\bibnamefont {Yu}},\ }\href {\doibase 10.1103/PhysRevE.93.053206}
  {\bibfield  {journal} {\bibinfo  {journal} {Phys. Rev. E}\ }\textbf {\bibinfo
  {volume} {93}},\ \bibinfo {pages} {053206} (\bibinfo {year}
  {2016})}\BibitemShut {NoStop}%
\bibitem [{\citenamefont {Rzazewski}\ \emph {et~al.}(2000)\citenamefont
  {Rzazewski}, \citenamefont {Plaja}, \citenamefont {Roso},\ and\ \citenamefont
  {von~der Linde}}]{RzazewskiJPB00}%
  \BibitemOpen
  \bibfield  {author} {\bibinfo {author} {\bibfnamefont {K.}~\bibnamefont
  {Rzazewski}}, \bibinfo {author} {\bibfnamefont {L.}~\bibnamefont {Plaja}},
  \bibinfo {author} {\bibfnamefont {L.}~\bibnamefont {Roso}}, \ and\ \bibinfo
  {author} {\bibfnamefont {D.}~\bibnamefont {von~der Linde}},\ }\href
  {http://stacks.iop.org/0953-4075/33/i=13/a=314} {\bibfield  {journal}
  {\bibinfo  {journal} {J. Phys. B: Atom. Mol. Opt. Phys.}\ }\textbf {\bibinfo
  {volume} {33}},\ \bibinfo {pages} {2549} (\bibinfo {year}
  {2000})}\BibitemShut {NoStop}%
\bibitem [{\citenamefont {Lavocat-Dubuis}\ and\ \citenamefont
  {Matte}(2009)}]{Lavocat-Dubuis_PRE_2009}%
  \BibitemOpen
  \bibfield  {author} {\bibinfo {author} {\bibfnamefont {X.}~\bibnamefont
  {Lavocat-Dubuis}}\ and\ \bibinfo {author} {\bibfnamefont {J.-P.}\
  \bibnamefont {Matte}},\ }\href {\doibase 10.1103/PhysRevE.80.055401}
  {\bibfield  {journal} {\bibinfo  {journal} {Phys. Rev. E}\ }\textbf {\bibinfo
  {volume} {80}},\ \bibinfo {pages} {055401} (\bibinfo {year}
  {2009})}\BibitemShut {NoStop}%
\bibitem [{\citenamefont {Cerchez}\ \emph {et~al.}(2013)\citenamefont
  {Cerchez}, \citenamefont {Giesecke}, \citenamefont {Peth}, \citenamefont
  {Toncian}, \citenamefont {Albertazzi}, \citenamefont {Fuchs}, \citenamefont
  {Willi},\ and\ \citenamefont {Toncian}}]{Cerchez_PRL2013}%
  \BibitemOpen
  \bibfield  {author} {\bibinfo {author} {\bibfnamefont {M.}~\bibnamefont
  {Cerchez}}, \bibinfo {author} {\bibfnamefont {A.~L.}\ \bibnamefont
  {Giesecke}}, \bibinfo {author} {\bibfnamefont {C.}~\bibnamefont {Peth}},
  \bibinfo {author} {\bibfnamefont {M.}~\bibnamefont {Toncian}}, \bibinfo
  {author} {\bibfnamefont {B.}~\bibnamefont {Albertazzi}}, \bibinfo {author}
  {\bibfnamefont {J.}~\bibnamefont {Fuchs}}, \bibinfo {author} {\bibfnamefont
  {O.}~\bibnamefont {Willi}}, \ and\ \bibinfo {author} {\bibfnamefont
  {T.}~\bibnamefont {Toncian}},\ }\href {\doibase
  10.1103/PhysRevLett.110.065003} {\bibfield  {journal} {\bibinfo  {journal}
  {Phys. Rev. Lett.}\ }\textbf {\bibinfo {volume} {110}},\ \bibinfo {pages}
  {065003} (\bibinfo {year} {2013})}\BibitemShut {NoStop}%
\bibitem [{\citenamefont {Ceccotti}\ \emph {et~al.}(2013)\citenamefont
  {Ceccotti}, \citenamefont {Floquet}, \citenamefont {Sgattoni}, \citenamefont
  {Bigongiari}, \citenamefont {Klimo}, \citenamefont {Raynaud}, \citenamefont
  {Riconda}, \citenamefont {Heron}, \citenamefont {Baffigi}, \citenamefont
  {Labate}, \citenamefont {Gizzi}, \citenamefont {Vassura}, \citenamefont
  {Fuchs}, \citenamefont {Passoni}, \citenamefont {Kv\ifmmode~\check{e}\else
  \v{e}\fi{}ton}, \citenamefont {Novotny}, \citenamefont {Possolt},
  \citenamefont {Prok\ifmmode~\mathring{u}\else \r{u}\fi{}pek}, \citenamefont
  {Pro\ifmmode~\check{s}\else \v{s}\fi{}ka}, \citenamefont
  {P\ifmmode~\check{s}\else \v{s}\fi{}ikal}, \citenamefont
  {\ifmmode~\check{S}\else \v{S}\fi{}tolcov\'a}, \citenamefont {Velyhan},
  \citenamefont {Bougeard}, \citenamefont {D'Oliveira}, \citenamefont
  {Tcherbakoff}, \citenamefont {R\'eau}, \citenamefont {Martin},\ and\
  \citenamefont {Macchi}}]{Ceccotti_PRL2013}%
  \BibitemOpen
  \bibfield  {author} {\bibinfo {author} {\bibfnamefont {T.}~\bibnamefont
  {Ceccotti}}, \bibinfo {author} {\bibfnamefont {V.}~\bibnamefont {Floquet}},
  \bibinfo {author} {\bibfnamefont {A.}~\bibnamefont {Sgattoni}}, \bibinfo
  {author} {\bibfnamefont {A.}~\bibnamefont {Bigongiari}}, \bibinfo {author}
  {\bibfnamefont {O.}~\bibnamefont {Klimo}}, \bibinfo {author} {\bibfnamefont
  {M.}~\bibnamefont {Raynaud}}, \bibinfo {author} {\bibfnamefont
  {C.}~\bibnamefont {Riconda}}, \bibinfo {author} {\bibfnamefont
  {A.}~\bibnamefont {Heron}}, \bibinfo {author} {\bibfnamefont
  {F.}~\bibnamefont {Baffigi}}, \bibinfo {author} {\bibfnamefont
  {L.}~\bibnamefont {Labate}}, \bibinfo {author} {\bibfnamefont {L.~A.}\
  \bibnamefont {Gizzi}}, \bibinfo {author} {\bibfnamefont {L.}~\bibnamefont
  {Vassura}}, \bibinfo {author} {\bibfnamefont {J.}~\bibnamefont {Fuchs}},
  \bibinfo {author} {\bibfnamefont {M.}~\bibnamefont {Passoni}}, \bibinfo
  {author} {\bibfnamefont {M.}~\bibnamefont {Kv\ifmmode~\check{e}\else
  \v{e}\fi{}ton}}, \bibinfo {author} {\bibfnamefont {F.}~\bibnamefont
  {Novotny}}, \bibinfo {author} {\bibfnamefont {M.}~\bibnamefont {Possolt}},
  \bibinfo {author} {\bibfnamefont {J.}~\bibnamefont
  {Prok\ifmmode~\mathring{u}\else \r{u}\fi{}pek}}, \bibinfo {author}
  {\bibfnamefont {J.}~\bibnamefont {Pro\ifmmode~\check{s}\else \v{s}\fi{}ka}},
  \bibinfo {author} {\bibfnamefont {J.}~\bibnamefont {P\ifmmode~\check{s}\else
  \v{s}\fi{}ikal}}, \bibinfo {author} {\bibfnamefont {L.}~\bibnamefont
  {\ifmmode~\check{S}\else \v{S}\fi{}tolcov\'a}}, \bibinfo {author}
  {\bibfnamefont {A.}~\bibnamefont {Velyhan}}, \bibinfo {author} {\bibfnamefont
  {M.}~\bibnamefont {Bougeard}}, \bibinfo {author} {\bibfnamefont
  {P.}~\bibnamefont {D'Oliveira}}, \bibinfo {author} {\bibfnamefont
  {O.}~\bibnamefont {Tcherbakoff}}, \bibinfo {author} {\bibfnamefont
  {F.}~\bibnamefont {R\'eau}}, \bibinfo {author} {\bibfnamefont
  {P.}~\bibnamefont {Martin}}, \ and\ \bibinfo {author} {\bibfnamefont
  {A.}~\bibnamefont {Macchi}},\ }\href {\doibase
  10.1103/PhysRevLett.111.185001} {\bibfield  {journal} {\bibinfo  {journal}
  {Phys. Rev. Lett.}\ }\textbf {\bibinfo {volume} {111}},\ \bibinfo {pages}
  {185001} (\bibinfo {year} {2013})}\BibitemShut {NoStop}%
\bibitem [{\citenamefont {Fedeli}\ \emph {et~al.}(2016)\citenamefont {Fedeli},
  \citenamefont {Sgattoni}, \citenamefont {Cantono}, \citenamefont {Garzella},
  \citenamefont {R\'eau}, \citenamefont {Prencipe}, \citenamefont {Passoni},
  \citenamefont {Raynaud}, \citenamefont {Kv\ifmmode \check{e}\else
  \v{e}\fi{}to\ifmmode~\check{n}\else \v{n}\fi{}}, \citenamefont {Proska},
  \citenamefont {Macchi},\ and\ \citenamefont {Ceccotti}}]{Fedeli_PRL2016}%
  \BibitemOpen
  \bibfield  {author} {\bibinfo {author} {\bibfnamefont {L.}~\bibnamefont
  {Fedeli}}, \bibinfo {author} {\bibfnamefont {A.}~\bibnamefont {Sgattoni}},
  \bibinfo {author} {\bibfnamefont {G.}~\bibnamefont {Cantono}}, \bibinfo
  {author} {\bibfnamefont {D.}~\bibnamefont {Garzella}}, \bibinfo {author}
  {\bibfnamefont {F.}~\bibnamefont {R\'eau}}, \bibinfo {author} {\bibfnamefont
  {I.}~\bibnamefont {Prencipe}}, \bibinfo {author} {\bibfnamefont
  {M.}~\bibnamefont {Passoni}}, \bibinfo {author} {\bibfnamefont
  {M.}~\bibnamefont {Raynaud}}, \bibinfo {author} {\bibfnamefont
  {M.}~\bibnamefont {Kv\ifmmode \check{e}\else
  \v{e}\fi{}to\ifmmode~\check{n}\else \v{n}\fi{}}}, \bibinfo {author}
  {\bibfnamefont {J.}~\bibnamefont {Proska}}, \bibinfo {author} {\bibfnamefont
  {A.}~\bibnamefont {Macchi}}, \ and\ \bibinfo {author} {\bibfnamefont
  {T.}~\bibnamefont {Ceccotti}},\ }\href {\doibase
  10.1103/PhysRevLett.116.015001} {\bibfield  {journal} {\bibinfo  {journal}
  {Phys. Rev. Lett.}\ }\textbf {\bibinfo {volume} {116}},\ \bibinfo {pages}
  {015001} (\bibinfo {year} {2016})}\BibitemShut {NoStop}%
\bibitem [{\citenamefont {Sgattoni}\ \emph {et~al.}(2016)\citenamefont
  {Sgattoni}, \citenamefont {Fedeli}, \citenamefont {Cantono}, \citenamefont
  {Ceccotti},\ and\ \citenamefont {Macchi}}]{SgattoPPCF}%
  \BibitemOpen
  \bibfield  {author} {\bibinfo {author} {\bibfnamefont {A.}~\bibnamefont
  {Sgattoni}}, \bibinfo {author} {\bibfnamefont {L.}~\bibnamefont {Fedeli}},
  \bibinfo {author} {\bibfnamefont {G.}~\bibnamefont {Cantono}}, \bibinfo
  {author} {\bibfnamefont {T.}~\bibnamefont {Ceccotti}}, \ and\ \bibinfo
  {author} {\bibfnamefont {A.}~\bibnamefont {Macchi}},\ }\href
  {http://stacks.iop.org/0741-3335/58/i=1/a=014004} {\bibfield  {journal}
  {\bibinfo  {journal} {Plasma Physics and Controlled Fusion}\ }\textbf
  {\bibinfo {volume} {58}},\ \bibinfo {pages} {014004} (\bibinfo {year}
  {2016})}\BibitemShut {NoStop}%
\bibitem [{\citenamefont {Kim}\ \emph {et~al.}(2008)\citenamefont {Kim},
  \citenamefont {Jin}, \citenamefont {Kim}, \citenamefont {Park}, \citenamefont
  {Kim},\ and\ \citenamefont {Kim}}]{kimN08}%
  \BibitemOpen
  \bibfield  {author} {\bibinfo {author} {\bibfnamefont {S.}~\bibnamefont
  {Kim}}, \bibinfo {author} {\bibfnamefont {J.}~\bibnamefont {Jin}}, \bibinfo
  {author} {\bibfnamefont {Y.-J.}\ \bibnamefont {Kim}}, \bibinfo {author}
  {\bibfnamefont {I.-Y.}\ \bibnamefont {Park}}, \bibinfo {author}
  {\bibfnamefont {Y.}~\bibnamefont {Kim}}, \ and\ \bibinfo {author}
  {\bibfnamefont {S.-W.}\ \bibnamefont {Kim}},\ }\href {\doibase
  10.1038/nature07012} {\bibfield  {journal} {\bibinfo  {journal} {Nature}\
  }\textbf {\bibinfo {volume} {453}},\ \bibinfo {pages} {757} (\bibinfo {year}
  {2008})}\BibitemShut {NoStop}%
\bibitem [{\citenamefont {Husakou}\ \emph {et~al.}(2011)\citenamefont
  {Husakou}, \citenamefont {Kelkensberg}, \citenamefont {Herrmann},\ and\
  \citenamefont {Vrakking}}]{HusakouOE11}%
  \BibitemOpen
  \bibfield  {author} {\bibinfo {author} {\bibfnamefont {A.}~\bibnamefont
  {Husakou}}, \bibinfo {author} {\bibfnamefont {F.}~\bibnamefont
  {Kelkensberg}}, \bibinfo {author} {\bibfnamefont {J.}~\bibnamefont
  {Herrmann}}, \ and\ \bibinfo {author} {\bibfnamefont {M.~J.~J.}\ \bibnamefont
  {Vrakking}},\ }\href {\doibase 10.1364/OE.19.025346} {\bibfield  {journal}
  {\bibinfo  {journal} {Opt. Express}\ }\textbf {\bibinfo {volume} {19}},\
  \bibinfo {pages} {25346} (\bibinfo {year} {2011})}\BibitemShut {NoStop}%
\bibitem [{\citenamefont {Park}\ \emph {et~al.}(2011)\citenamefont {Park},
  \citenamefont {Kim}, \citenamefont {Choi}, \citenamefont {Lee}, \citenamefont
  {Kim}, \citenamefont {Kling}, \citenamefont {Stockman},\ and\ \citenamefont
  {Kim}}]{parkN11}%
  \BibitemOpen
  \bibfield  {author} {\bibinfo {author} {\bibfnamefont {I.-Y.}\ \bibnamefont
  {Park}}, \bibinfo {author} {\bibfnamefont {S.}~\bibnamefont {Kim}}, \bibinfo
  {author} {\bibfnamefont {J.}~\bibnamefont {Choi}}, \bibinfo {author}
  {\bibfnamefont {D.-H.}\ \bibnamefont {Lee}}, \bibinfo {author} {\bibfnamefont
  {Y.-J.}\ \bibnamefont {Kim}}, \bibinfo {author} {\bibfnamefont {M.~F.}\
  \bibnamefont {Kling}}, \bibinfo {author} {\bibfnamefont {M.~I.}\ \bibnamefont
  {Stockman}}, \ and\ \bibinfo {author} {\bibfnamefont {S.-W.}\ \bibnamefont
  {Kim}},\ }\href {\doibase 10.1038/nphoton.2011.258} {\bibfield  {journal}
  {\bibinfo  {journal} {Nature Phot.}\ }\textbf {\bibinfo {volume} {5}},\
  \bibinfo {pages} {677} (\bibinfo {year} {2011})}\BibitemShut {NoStop}%
\bibitem [{\citenamefont {Sivis}\ \emph {et~al.}(2013)\citenamefont {Sivis},
  \citenamefont {Duwe}, \citenamefont {Abel},\ and\ \citenamefont
  {Ropers}}]{sivisNP13}%
  \BibitemOpen
  \bibfield  {author} {\bibinfo {author} {\bibfnamefont {M.}~\bibnamefont
  {Sivis}}, \bibinfo {author} {\bibfnamefont {M.}~\bibnamefont {Duwe}},
  \bibinfo {author} {\bibfnamefont {B.}~\bibnamefont {Abel}}, \ and\ \bibinfo
  {author} {\bibfnamefont {C.}~\bibnamefont {Ropers}},\ }\href {\doibase
  10.1038/nphys2590} {\bibfield  {journal} {\bibinfo  {journal} {Nature Phys.}\
  }\textbf {\bibinfo {volume} {9}},\ \bibinfo {pages} {304} (\bibinfo {year}
  {2013})}\BibitemShut {NoStop}%
\bibitem [{\citenamefont {Hurst}\ \emph {et~al.}(2014)\citenamefont {Hurst},
  \citenamefont {Haas}, \citenamefont {Manfredi},\ and\ \citenamefont
  {Hervieux}}]{hurstPRB14}%
  \BibitemOpen
  \bibfield  {author} {\bibinfo {author} {\bibfnamefont {J.}~\bibnamefont
  {Hurst}}, \bibinfo {author} {\bibfnamefont {F.}~\bibnamefont {Haas}},
  \bibinfo {author} {\bibfnamefont {G.}~\bibnamefont {Manfredi}}, \ and\
  \bibinfo {author} {\bibfnamefont {P.-A.}\ \bibnamefont {Hervieux}},\ }\href
  {\doibase 10.1103/PhysRevB.89.161111} {\bibfield  {journal} {\bibinfo
  {journal} {Phys. Rev. B}\ }\textbf {\bibinfo {volume} {89}},\ \bibinfo
  {pages} {161111} (\bibinfo {year} {2014})}\BibitemShut {NoStop}%
\bibitem [{\citenamefont {Han}\ \emph {et~al.}(2016)\citenamefont {Han},
  \citenamefont {Kim}, \citenamefont {Kim}, \citenamefont {Kim}, \citenamefont
  {Kim}, \citenamefont {Park},\ and\ \citenamefont {Kim}}]{hanNC16}%
  \BibitemOpen
  \bibfield  {author} {\bibinfo {author} {\bibfnamefont {S.}~\bibnamefont
  {Han}}, \bibinfo {author} {\bibfnamefont {H.}~\bibnamefont {Kim}}, \bibinfo
  {author} {\bibfnamefont {Y.~W.}\ \bibnamefont {Kim}}, \bibinfo {author}
  {\bibfnamefont {Y.-J.}\ \bibnamefont {Kim}}, \bibinfo {author} {\bibfnamefont
  {S.}~\bibnamefont {Kim}}, \bibinfo {author} {\bibfnamefont {I.-Y.}\
  \bibnamefont {Park}}, \ and\ \bibinfo {author} {\bibfnamefont {S.-W.}\
  \bibnamefont {Kim}},\ }\href {\doibase 10.1038/ncomms13105} {\bibfield
  {journal} {\bibinfo  {journal} {Nature Communications}\ }\textbf {\bibinfo
  {volume} {7}},\ \bibinfo {pages} {13105} (\bibinfo {year}
  {2016})}\BibitemShut {NoStop}%
\bibitem [{\citenamefont {Maier}(2007)}]{bookMaier2007}%
  \BibitemOpen
  \bibfield  {author} {\bibinfo {author} {\bibfnamefont {S.~A.}\ \bibnamefont
  {Maier}},\ }\href@noop {} {\emph {\bibinfo {title} {Plasmonics: fundamentals
  and applications}}}\ (\bibinfo  {publisher} {Springer Science \& Business
  Media},\ \bibinfo {year} {2007})\BibitemShut {NoStop}%
\bibitem [{\citenamefont {Landau}\ and\ \citenamefont
  {Lifshitz}(1960)}]{landau8}%
  \BibitemOpen
  \bibfield  {author} {\bibinfo {author} {\bibfnamefont {L.~D.}\ \bibnamefont
  {Landau}}\ and\ \bibinfo {author} {\bibfnamefont {E.~M.}\ \bibnamefont
  {Lifshitz}},\ }\href@noop {} {\emph {\bibinfo {title} {Electrodynamics of
  Continuous Media}}}\ (\bibinfo  {publisher} {Pergamon},\ \bibinfo {year}
  {1960})\BibitemShut {NoStop}%
\bibitem [{\citenamefont {Lavocat-Dubuis}\ and\ \citenamefont
  {Matte}(2010)}]{Lavocat-Dubuis_PoP_2010}%
  \BibitemOpen
  \bibfield  {author} {\bibinfo {author} {\bibfnamefont {X.}~\bibnamefont
  {Lavocat-Dubuis}}\ and\ \bibinfo {author} {\bibfnamefont {J.-P.}\
  \bibnamefont {Matte}},\ }\href {\doibase 10.1063/1.3485119} {\bibfield
  {journal} {\bibinfo  {journal} {Phys. Plasmas}\ }\textbf {\bibinfo {volume}
  {17}},\ \bibinfo {eid} {093105} (\bibinfo {year} {2010}),\
  10.1063/1.3485119}\BibitemShut {NoStop}%
\bibitem [{\citenamefont {Dromey}\ \emph {et~al.}(2004)\citenamefont {Dromey},
  \citenamefont {Kar}, \citenamefont {Zepf},\ and\ \citenamefont
  {Foster}}]{Dromey_RSI_2004}%
  \BibitemOpen
  \bibfield  {author} {\bibinfo {author} {\bibfnamefont {B.}~\bibnamefont
  {Dromey}}, \bibinfo {author} {\bibfnamefont {S.}~\bibnamefont {Kar}},
  \bibinfo {author} {\bibfnamefont {M.}~\bibnamefont {Zepf}}, \ and\ \bibinfo
  {author} {\bibfnamefont {P.}~\bibnamefont {Foster}},\ }\href {\doibase
  http://dx.doi.org/10.1063/1.1646737} {\bibfield  {journal} {\bibinfo
  {journal} {Review of Scientific Instruments}\ }\textbf {\bibinfo {volume}
  {75}},\ \bibinfo {pages} {645} (\bibinfo {year} {2004})}\BibitemShut
  {NoStop}%
\bibitem [{\citenamefont {Levy}\ \emph {et~al.}(2007)\citenamefont {Levy},
  \citenamefont {Ceccotti}, \citenamefont {D'Oliveira}, \citenamefont
  {R\'{e}au}, \citenamefont {Perdrix}, \citenamefont {Qu\'{e}r\'{e}},
  \citenamefont {Monot}, \citenamefont {Bougeard}, \citenamefont {Lagadec},
  \citenamefont {Martin}, \citenamefont {Geindre},\ and\ \citenamefont
  {Audebert}}]{Levy_OptLett2007}%
  \BibitemOpen
  \bibfield  {author} {\bibinfo {author} {\bibfnamefont {A.}~\bibnamefont
  {Levy}}, \bibinfo {author} {\bibfnamefont {T.}~\bibnamefont {Ceccotti}},
  \bibinfo {author} {\bibfnamefont {P.}~\bibnamefont {D'Oliveira}}, \bibinfo
  {author} {\bibfnamefont {F.}~\bibnamefont {R\'{e}au}}, \bibinfo {author}
  {\bibfnamefont {M.}~\bibnamefont {Perdrix}}, \bibinfo {author} {\bibfnamefont
  {F.}~\bibnamefont {Qu\'{e}r\'{e}}}, \bibinfo {author} {\bibfnamefont
  {P.}~\bibnamefont {Monot}}, \bibinfo {author} {\bibfnamefont
  {M.}~\bibnamefont {Bougeard}}, \bibinfo {author} {\bibfnamefont
  {H.}~\bibnamefont {Lagadec}}, \bibinfo {author} {\bibfnamefont
  {P.}~\bibnamefont {Martin}}, \bibinfo {author} {\bibfnamefont
  {P.}~\bibnamefont {Geindre}}, \ and\ \bibinfo {author} {\bibfnamefont
  {P.}~\bibnamefont {Audebert}},\ }\href {\doibase 10.1364/OL.32.000310}
  {\bibfield  {journal} {\bibinfo  {journal} {Opt. Lett.}\ }\textbf {\bibinfo
  {volume} {32}},\ \bibinfo {pages} {310} (\bibinfo {year} {2007})}\BibitemShut
  {NoStop}%
\bibitem [{\citenamefont {Kapteyn}\ \emph {et~al.}(1991)\citenamefont
  {Kapteyn}, \citenamefont {Szoke}, \citenamefont {Falcone},\ and\
  \citenamefont {Murnane}}]{Kapten_OptLett1991}%
  \BibitemOpen
  \bibfield  {author} {\bibinfo {author} {\bibfnamefont {H.~C.}\ \bibnamefont
  {Kapteyn}}, \bibinfo {author} {\bibfnamefont {A.}~\bibnamefont {Szoke}},
  \bibinfo {author} {\bibfnamefont {R.~W.}\ \bibnamefont {Falcone}}, \ and\
  \bibinfo {author} {\bibfnamefont {M.~M.}\ \bibnamefont {Murnane}},\ }\href
  {\doibase 10.1364/OL.16.000490} {\bibfield  {journal} {\bibinfo  {journal}
  {Opt. Lett.}\ }\textbf {\bibinfo {volume} {16}},\ \bibinfo {pages} {490}
  (\bibinfo {year} {1991})}\BibitemShut {NoStop}%
\bibitem [{\citenamefont {Thaury}\ \emph {et~al.}(2008)\citenamefont {Thaury},
  \citenamefont {Qu\'{e}r\'{e}}, \citenamefont {Geindre}, \citenamefont {Levy},
  \citenamefont {Ceccotti}, \citenamefont {Monot}, \citenamefont {Bougeard},
  \citenamefont {Reau}, \citenamefont {d'Oliveira}, \citenamefont {Audebert},
  \citenamefont {Marjoribanks},\ and\ \citenamefont
  {Martin}}]{Thaury_NatPhys2008}%
  \BibitemOpen
  \bibfield  {author} {\bibinfo {author} {\bibfnamefont {C.}~\bibnamefont
  {Thaury}}, \bibinfo {author} {\bibfnamefont {F.}~\bibnamefont
  {Qu\'{e}r\'{e}}}, \bibinfo {author} {\bibfnamefont {J.-P.}\ \bibnamefont
  {Geindre}}, \bibinfo {author} {\bibfnamefont {A.}~\bibnamefont {Levy}},
  \bibinfo {author} {\bibfnamefont {T.}~\bibnamefont {Ceccotti}}, \bibinfo
  {author} {\bibfnamefont {P.}~\bibnamefont {Monot}}, \bibinfo {author}
  {\bibfnamefont {M.}~\bibnamefont {Bougeard}}, \bibinfo {author}
  {\bibfnamefont {F.}~\bibnamefont {Reau}}, \bibinfo {author} {\bibfnamefont
  {P.}~\bibnamefont {d'Oliveira}}, \bibinfo {author} {\bibfnamefont
  {P.}~\bibnamefont {Audebert}}, \bibinfo {author} {\bibfnamefont
  {R.}~\bibnamefont {Marjoribanks}}, \ and\ \bibinfo {author} {\bibfnamefont
  {P.}~\bibnamefont {Martin}},\ }\href@noop {} {\bibfield  {journal} {\bibinfo
  {journal} {Nature Phys.}\ }\textbf {\bibinfo {volume} {6}} (\bibinfo {year}
  {2008})}\BibitemShut {NoStop}%
\bibitem [{\citenamefont {Sgattoni}, \citenamefont {Fedeli},\ and\
  \citenamefont {Sinigardi}()}]{piccante}%
  \BibitemOpen
  \bibfield  {author} {\bibinfo {author} {\bibfnamefont {A.}~\bibnamefont
  {Sgattoni}}, \bibinfo {author} {\bibfnamefont {L.}~\bibnamefont {Fedeli}}, \
  and\ \bibinfo {author} {\bibfnamefont {S.}~\bibnamefont {Sinigardi}},\
  }\href@noop {} {\enquote {\bibinfo {title} {Piccante: an open-source, fully
  relativistic, massively parallel particle-in-cell code},}\ }\bibinfo
  {howpublished} {http://aladyn.github.io/piccante/}\BibitemShut {NoStop}%
\bibitem [{\citenamefont {{Sgattoni}}\ \emph {et~al.}(2015)\citenamefont
  {{Sgattoni}}, \citenamefont {{Fedeli}}, \citenamefont {{Sinigardi}},
  \citenamefont {{Marocchino}}, \citenamefont {{Macchi}}, \citenamefont
  {{Weinberg}},\ and\ \citenamefont {{Karmakar}}}]{arxiv_piccante}%
  \BibitemOpen
  \bibfield  {author} {\bibinfo {author} {\bibfnamefont {A.}~\bibnamefont
  {{Sgattoni}}}, \bibinfo {author} {\bibfnamefont {L.}~\bibnamefont
  {{Fedeli}}}, \bibinfo {author} {\bibfnamefont {S.}~\bibnamefont
  {{Sinigardi}}}, \bibinfo {author} {\bibfnamefont {A.}~\bibnamefont
  {{Marocchino}}}, \bibinfo {author} {\bibfnamefont {A.}~\bibnamefont
  {{Macchi}}}, \bibinfo {author} {\bibfnamefont {V.}~\bibnamefont
  {{Weinberg}}}, \ and\ \bibinfo {author} {\bibfnamefont {A.}~\bibnamefont
  {{Karmakar}}},\ }\href {http://arxiv.org/abs/1503.02464} {\bibfield
  {journal} {\bibinfo  {journal} {ArXiv e-prints}\ } (\bibinfo {year}
  {2015})},\ \Eprint {http://arxiv.org/abs/1503.02464} {arXiv:1503.02464
  [cs.DC]} \BibitemShut {NoStop}%
\bibitem [{\citenamefont {Danson}\ \emph {et~al.}(2015)\citenamefont {Danson},
  \citenamefont {Hillier}, \citenamefont {Hopps},\ and\ \citenamefont
  {Neely}}]{HPL:9533946}%
  \BibitemOpen
  \bibfield  {author} {\bibinfo {author} {\bibfnamefont {C.}~\bibnamefont
  {Danson}}, \bibinfo {author} {\bibfnamefont {D.}~\bibnamefont {Hillier}},
  \bibinfo {author} {\bibfnamefont {N.}~\bibnamefont {Hopps}}, \ and\ \bibinfo
  {author} {\bibfnamefont {D.}~\bibnamefont {Neely}},\ }\href {\doibase
  10.1017/hpl.2014.52} {\bibfield  {journal} {\bibinfo  {journal} {High Power
  Laser Science and Engineering}\ }\textbf {\bibinfo {volume} {3}},\ \bibinfo
  {pages} {e3} (\bibinfo {year} {2015})}\BibitemShut {NoStop}%
\bibitem [{\citenamefont {Bigongiari}\ \emph {et~al.}(2011)\citenamefont
  {Bigongiari}, \citenamefont {Raynaud}, \citenamefont {Riconda}, \citenamefont
  {H\'eron},\ and\ \citenamefont {Macchi}}]{BigongiariPoP2011}%
  \BibitemOpen
  \bibfield  {author} {\bibinfo {author} {\bibfnamefont {A.}~\bibnamefont
  {Bigongiari}}, \bibinfo {author} {\bibfnamefont {M.}~\bibnamefont {Raynaud}},
  \bibinfo {author} {\bibfnamefont {C.}~\bibnamefont {Riconda}}, \bibinfo
  {author} {\bibfnamefont {A.}~\bibnamefont {H\'eron}}, \ and\ \bibinfo
  {author} {\bibfnamefont {A.}~\bibnamefont {Macchi}},\ }\href
  {http://scitation.aip.org/content/aip/journal/pop/18/10/10.1063/1.3646520}
  {\bibfield  {journal} {\bibinfo  {journal} {Phys. Plasmas}\ }\textbf
  {\bibinfo {volume} {18}},\ \bibinfo {eid} {102701} (\bibinfo {year}
  {2011})}\BibitemShut {NoStop}%
\bibitem [{\citenamefont {Monchoc\'e}\ \emph {et~al.}(2014)\citenamefont
  {Monchoc\'e}, \citenamefont {Kahaly}, \citenamefont {Leblanc}, \citenamefont
  {Videau}, \citenamefont {Combis}, \citenamefont {R\'eau}, \citenamefont
  {Garzella}, \citenamefont {D'Oliveira}, \citenamefont {Martin},\ and\
  \citenamefont {Qu\'er\'e}}]{monchocePRL14}%
  \BibitemOpen
  \bibfield  {author} {\bibinfo {author} {\bibfnamefont {S.}~\bibnamefont
  {Monchoc\'e}}, \bibinfo {author} {\bibfnamefont {S.}~\bibnamefont {Kahaly}},
  \bibinfo {author} {\bibfnamefont {A.}~\bibnamefont {Leblanc}}, \bibinfo
  {author} {\bibfnamefont {L.}~\bibnamefont {Videau}}, \bibinfo {author}
  {\bibfnamefont {P.}~\bibnamefont {Combis}}, \bibinfo {author} {\bibfnamefont
  {F.}~\bibnamefont {R\'eau}}, \bibinfo {author} {\bibfnamefont
  {D.}~\bibnamefont {Garzella}}, \bibinfo {author} {\bibfnamefont
  {P.}~\bibnamefont {D'Oliveira}}, \bibinfo {author} {\bibfnamefont
  {P.}~\bibnamefont {Martin}}, \ and\ \bibinfo {author} {\bibfnamefont
  {F.}~\bibnamefont {Qu\'er\'e}},\ }\href {\doibase
  10.1103/PhysRevLett.112.145008} {\bibfield  {journal} {\bibinfo  {journal}
  {Phys. Rev. Lett.}\ }\textbf {\bibinfo {volume} {112}},\ \bibinfo {pages}
  {145008} (\bibinfo {year} {2014})}\BibitemShut {NoStop}%
\end{thebibliography}
\end{document}